\documentclass[nofootinbib,amsmath,amssymb,aps,floatfix,notitlepage]{revtex4-1}

\pdfoutput=1

\usepackage{graphicx}
\usepackage{dcolumn}
\usepackage{bm}

\usepackage{amssymb}
\usepackage{amsmath}
\usepackage{epstopdf}
\usepackage{xcolor}
\usepackage{mathtools}
\usepackage{comment}

\def\hhref#1{\href{http://arxiv.org/abs/#1}{arXiv:#1}} 
\usepackage{hyperref}

\def\hhref#1{\href{http://arxiv.org/abs/#1}{arXiv:#1}} 
\begin{document}  

\title{Resurgent Extrapolation: Rebuilding a Function from Asymptotic Data. Painlev\'e I}

\author{Ovidiu Costin$^{1}$ and Gerald~V.~Dunne$^{2}$}

\affiliation{
$^1$ Department of Mathematics, The Ohio State University, Columbus, OH 43210\\
$^2$Department of Physics, University of Connecticut, Storrs, CT 06269}

\begin{abstract}
Extrapolation is a  generic problem in physics and mathematics: how to use asymptotic data in one parametric regime to learn about the behavior of a function in another parametric regime. For example: extending weak coupling expansions to strong coupling, or high temperature expansions to low temperature, or vice versa. Such extrapolations are particularly interesting in systems possessing dualities. Here we study numerical procedures for performing such an extrapolation, combining ideas from resurgent asymptotics with well-known techniques of Borel summation, Pad\'e approximants and conformal mapping. We illustrate the method with the concrete example of the Painlev\'e I equation, which has applications in many branches of physics and mathematics. Starting solely with a finite number of coefficients from asymptotic data at infinity on the positive real line, we obtain a high precision extrapolation of the function throughout the complex plane, even across the phase transition into the pole region. The precision far exceeds that of state-of-the-art numerical integration methods along the real axis. The methods used are both elementary and general, not relying on Painlev\'e integrability properties, and so are applicable to a wide class of extrapolation problems.

\end{abstract}

\maketitle

\section{Introduction}

Since exact solutions are rare in physics, asymptotic analysis is a common and powerful method for studying physical systems at extreme values of the relevant parameters, such as couplings, or masses, or temperature, or density, etc. For non-trivial physical systems, it is usually only possible to generate a {\it finite} number of terms in such an expansion. This raises the important question: how much information about the function being computed is encoded in this finite set of terms, generated in one particular asymptotic region? One class of such questions is the "central connection problem": how can one extrapolate from strong coupling to weak coupling, or high temperature to low temperature, or high density to low density? This may enable access to the more difficult regime of {\it intermediate} values (neither small nor large) of coupling, or temperature, or density.  Another class of questions involves extrapolation of the physical parameter from {\it real} values to {\it complex} values, or {\it vice versa}, a paradigm of which is the Lee-Yang-Fisher characterization of phase transitions in terms of complex zeros of the physical partition function \cite{lee-yang,fisher-zeros}. Motivated by these considerations, in this paper we study numerical methods to extrapolate finite-order perturbative expansions obtained in one parametric regime, at infinity in our chosen variable, down to zero and then throughout the entire complex plane. Extrapolation is of course a well-studied problem \cite{fisher-series,carl-book,zinn-qft,kleinert,caliceti}, and our new contribution is to demonstrate how and why ideas from resurgent asymptotics 
 \cite{dingle,ecalle,costin-book,sauzin,Aniceto:2013fka,Dorigoni:2014hea,gokce} significantly improve the numerical reach of such extrapolations. A subsequent paper \cite{cd-2} uses resurgent asymptotics to provide precise analytic estimates of the amount of information that can be extracted from a given number of terms, also based on the precision to which these terms are known.
An intuitive explanation of why resurgence  provides an advantage is that resurgent functions have an orderly structure in the Borel plane, which suggests that it may be possible to characterize or parametrize a resurgent function with a relatively small number of coefficients, if the extrapolation method is able to encode  the resurgent structure in an efficient way. In this sense, our motivation also includes the possibility to extend conventional  numerical algorithms to incorporate  resurgent structure explicitly.  

In this paper, we study the extrapolation of a particular concrete example, the Painlev\'e I equation. This choice is made for several reasons: (i) the Painlev\'e non-linear differential equations have many interesting applications in physics and mathematics \cite{clarkson,mccoy-wu,mason,gromak,fokas,forrester-book,tracy-widom,bonelli}, with the Painlev\'e I equation being of particular interest for matrix models and 2d quantum gravity \cite{DiFrancesco:1993cyw,Fokas:1990wb,silvestrov,david,marino-matrix}; (ii) the Painlev\'e I solutions have non-trivial analytic structure in the complex plane \cite{costin-odes,costin-dmj,costin-costin-inventiones,costin-costin-huang,boutroux,kapaev,kitaev,joshi,takei,dubrovin,novokshenov,Garoufalidis:2010ya,Aniceto:2011nu,costin-huang-tanveer,Lisovyy:2016qig,Iwaki:2015xnr}, which illustrates extrapolation from large to small parameter, and also Stokes phase transitions in the complex plane; (iii) the known analytic structure of Painlev\'e solutions permits high-precision diagnostic tests of the quality of our extrapolations. We stress that even though the Painlev\'e equations are of course very special, since they are integrable in the sense of Painlev\'e \cite{ince}, the methods we use do not rely on this integrability. 
The class of resurgent problems is much larger than that of integrable problems, and indeed resurgence applies to all ``natural problems'' \cite{ecalle}, for example those based on differential, or difference, or integral, equations, and so resurgent extrapolation methods are expected to have  broad applicability in physics.

Section II explains how to generate our input "perturbative data" for the Painlev\'e I equation, and outlines the goals of our extrapolation. Section III reviews the well-known ideas of Borel summation, Pad\'e approximants and conformal mapping applied to extrapolation along the real axis, as applied to Painlev\'e I. We show that conformal mapping in the Borel plane significantly improves Pad\'e-Borel extrapolation, and in Section IV we argue that this is because the conformal mapping reveals and encodes the underlying resurgent structure of the function being computed. In Section V we introduce a re-expansion method, based on Pad\'e analysis in the physical plane, using our high-precision extrapolation along the real axis, which yields an extrapolation throughout the complex plane, crossing  the phase transition into the {\it tritronque\'e} pole region. Remarkably, only a modest amount of starting "perturbative data" is required in order to obtain non-trivial information about this phase transition and the structure of the pole region, which we confirm by comparing with known connection formulae and asymptotic estimates of pole locations.

\section{Painlev\'e I Equation: ``Perturbative'' Asymptotic Input Data}
\label{sec:data}

The Painlev\'e I equation (referred to below as PI)  in standard form reads \cite{nist-painleve,joshi,dubrovin}:
\begin{eqnarray}
y''(x)=6\, y^2(x) - x
\label{eq:p1}
\end{eqnarray}
Seeking a smooth real solution at large positive $x$ leads to an asymptotic expansion of the form
\begin{eqnarray}
y(x)\sim -\sqrt{\frac{x}{6}}\sum_{n=0}^\infty a_n\, \left(\frac{30}{\left(24 \, x\right)^{5/4}}\right)^{2n}
\qquad, \qquad x\to+\infty
\label{eq:largex}
\end{eqnarray}
where $a_0\equiv 1$, and with this choice of normalization all the expansion coefficients $a_n$ are rational numbers [see (\ref{eq:recursion})]. We define  the natural "\'Ecalle time" variable
\begin{eqnarray}
t\equiv \frac{\left(24\, x\right)^{5/4}}{30}
\label{eq:ecallet}
\end{eqnarray}
It is convenient to extract the overall square root behavior and define the function $h(t)$ by
\begin{eqnarray}
y(x)\equiv -\sqrt{\frac{x}{6}}\bigg(1+h(t)\bigg)
\label{eq:yh}
\end{eqnarray}
In terms of $h(t)$ the PI equation (\ref{eq:p1}) becomes:
\begin{eqnarray}
\ddot{h}+\frac{1}{t}\,\dot{h}+h\left(1+\frac{1}{2} h\right) -\frac{4}{25\, t^2} \left(1+h\right)=0
\label{eq:p1h}
\end{eqnarray}
where the overdot symbol denotes $\frac{d}{dt}$.
The  $x\to+\infty$ asymptotic expansion (\ref{eq:largex})  for $y(x)$ becomes a  $t\to+\infty$ asymptotic expansion for $h(t)$: 
\begin{eqnarray}
h(t)\sim \sum_{n=1}^\infty \frac{a_n}{t^{2n}}
\qquad, \qquad t\to+\infty
\label{eq:h}
\end{eqnarray}
where the $a_n$ are the same numerical coefficients as in (\ref{eq:largex}).
Our strategy will be to analyze the function $h(t)$, and then use it to reconstruct the physical PI solution $y(x)$ via the definitions  
(\ref{eq:ecallet})-(\ref{eq:yh}). 

The rational expansion coefficients $a_n$ in (\ref{eq:largex}) and (\ref{eq:h})  are generated from the recursion relation
\begin{eqnarray}
a_n&=&-4(n-1)^2a_{n-1}-\frac{1}{2}\sum_{m=2}^{n-2} a_m\, a_{n-m}\qquad, \quad n\geq 3 
\nonumber\\
&& a_1=\frac{4}{25}\qquad , \qquad a_2=-\frac{392}{625}
\label{eq:recursion}
\end{eqnarray}
We take a certain number, $N$, of these coefficients as our input ``perturbative data'':
\begin{eqnarray}
\text{perturbative input data}&=& \left\{a_1, a_2, \dots, a_{N} \right\} \nonumber\\
&=&\{\frac{4}{25}, -\frac{392}{625}, \frac{6\,272}{625}, -\frac{141\,196\,832}{390\,625}, 
\frac{9\,039\,055\,872}{390\,625}, \dots, a_{N} \}
\label{eq:input}
\end{eqnarray} 
Our goal is to learn as much as possible about the function $y(x)$, starting solely from the perturbative input data (\ref{eq:input}). The Painlev\'e I equation (\ref{eq:p1}) is used in our analysis to generate the input data [i.e., the coefficients $a_n$ in (\ref{eq:input})], and later as a diagnostic tool to test the level of precision of the resulting extrapolation. 

We have the following specific technical goals:
\begin{enumerate}
\item
Extrapolation of the formal "perturbative" expansion at $x=+\infty$ in (\ref{eq:largex}), starting with a {\it finite} number of terms, along the real axis all the way down to $x=0$. This is the classical central connection problem for PI, for which there is no known closed-form solution. This is an analogue of determining strong-coupling behavior from weak-coupling asymptotics, or vice versa.

\item
Extrapolation of the function $y(x)$ into the complex $x$ plane, once again starting just from the formal "perturbative" expansion at $x=+\infty$ in (\ref{eq:largex}), with a {\it finite} number of terms. At this stage, resurgent asymptotics begins to play a crucial role, both in terms of increased numerical precision, and also in terms of how much of the complex $x$ plane can be explored accurately.

\item
As a first step of the extrapolation into the complex plane, we show that the perturbative large  $x$ data in (\ref{eq:input}) permits a remarkably high-precision extraction of the PI Stokes constant, which is known analytically from isomonodromy methods \cite{fokas,kapaev} and also from resurgent asymptotics \cite{costin-costin-inventiones,costin-costin-huang}. This enables probing of Stokes transitions, and access to higher Riemann sheets, purely from the asymptotic data on the positive real line. This is an analogue of determining non-perturbative effects from perturbative data.

\item
Exploration of the transition into the PI pole region. It is well known that while the general solution to PI has poles throughout the complex $x$ plane, distributed asymptotically according to those of an associated Weierstrass elliptic function \cite{boutroux,kapaev,kitaev,joshi}, the formal expansion (\ref{eq:largex}) defines Boutroux's {\it tritronqu\'ee} solution to PI, which has poles in the complex $x$ plane only in a wedge region of opening angle $\frac{2\pi}{5}$ centered on the negative real $x$ axis \cite{dubrovin,costin-huang-tanveer}. We seek to explore this pole region numerically, mapping out its distribution and properties, once again starting just from the formal "perturbative" expansion at $x=+\infty$ in (\ref{eq:largex}), with a {\it finite} number of input coefficients.
This is an analogue of probing a phase transition using perturbative expansion data generated from a point well away from the transition region. The behavior of the PI function $y(x)$ changes radically as one crosses into the pole region, and its asymptotic trans-series expansion undergoes a dramatic rearrangement: the formal expansion (\ref{eq:largex}) is completely different from the form of the function in the pole region: see Eq. (\ref{eq:general}). We seek to learn as much as possible about this transition  
from the finite perturbative input data in (\ref{eq:input}).

\end{enumerate}

In our numerical extrapolation procedure we combine and compare several standard methods, such as Borel summation, Pad\'e approximants and conformal mapping \cite{fisher-series,carl-book,zinn-qft,kleinert,caliceti}. We  incorporate ideas from resurgent asymptotics, with the goal of developing new extrapolation methods of increased precision and enlarged region of validity. Resurgence explains why conformal mapping is such a powerful step in this analysis. The resurgence of the Painlev\'e I equation is well established by general theorems and explicit computations
 \cite{Garoufalidis:2010ya,Aniceto:2011nu,costin-huang-tanveer,costin-costin-inventiones,costin-costin-huang,marino-matrix,costin-odes,costin-dmj,Iwaki:2015xnr}, and here our numerical analysis provides further numerical evidence of these features.

\section{Extrapolation Along the Real Axis}
\label{sec:real-axis}

In this Section we extrapolate the PI solution $y(x)$, starting with a finite number of terms in its formal asymptotic expansion (\ref{eq:largex}) at $x=+\infty$, along the positive  real axis down to $x=0$, using various combinations of standard techniques, combined with some new improvements motivated by resurgent asymptotics. We compare the increased level of precision as the extrapolation method is refined. We perform our extrapolation directly on the formal asymptotic expansion (\ref{eq:h}) of the function $h(t)$ defined in (\ref{eq:yh}),  and then map back to the physical PI solution $y(x)$ using the definitions (\ref{eq:largex})--(\ref{eq:yh}).

\subsection{Borel Transform}
\label{sec:borel}

The first observation is that the perturbative coefficients $a_n$ in (\ref{eq:largex}) and (\ref{eq:h}), generated from the recursion formula (\ref{eq:recursion}), alternate in sign and grow factorially fast in magnitude. The alternating sign property is directly correlated with the choice of overall sign in  (\ref{eq:largex}) \cite{kapaev,joshi,novokshenov}. With just 10 perturbative coefficients $\{a_1, a_2, \dots a_{10}\}$, straightforward Richardson extrapolation \cite{carl-book} identifies the leading rate of growth as
\begin{eqnarray}
a_n\sim (0.1967...) (-1)^{n+1} \Gamma\left(2n-\frac{1}{2}\right)+\dots\qquad, \qquad n\to\infty
\label{eq:factorial}
\end{eqnarray}
with the overall coefficient correct to 4 digits. Using 50 input coefficients permits a high-precision numerical identification of the overall coefficient, as well as extraction of subleading corrections:
\begin{eqnarray}
a_n&\sim& \frac{1}{\pi}\sqrt{\frac{6}{5\pi}} (-1)^{n+1} \Gamma\left(2n-\frac{1}{2}\right)\left(1-\frac{\frac{1}{8}}{\left(2n-\frac{3}{2}\right)}+\frac{\frac{9}{128}}{\left(2n-\frac{3}{2}\right)\left(2n-\frac{5}{2}\right)} \right.
\nonumber\\
&&\left. -
\frac{\frac{341\,329}{1\,920\,000}}{\left(2n-\frac{3}{2}\right)\left(2n-\frac{5}{2}\right)\left(2n-\frac{7}{2}\right)}
+\dots\right)
\qquad, \qquad n\to\infty
\label{eq:factorial1}
\end{eqnarray}
One can verify that the coefficients of the subleading corrections coincide with the low order coefficients of the expansion about the  first exponential term in the trans-series expansion of the solution, as implied by general results of resurgent asymptotics \cite{dingle,ecalle,berry-howls,sauzin,costin-book,Aniceto:2013fka,Dorigoni:2014hea,gokce}.

The factorial growth in (\ref{eq:factorial})-(\ref{eq:factorial1}) implies that the expansions (\ref{eq:largex}) and (\ref{eq:h}) are formal divergent series, and therefore a natural next step is to define the Borel transform \cite{carl-book,costin-book} as the inverse Laplace transform of $h(t)$:
\begin{eqnarray}
\text{Borel transform:} \qquad  \mathcal B [h](p)\equiv \sum_{n=1}^\infty \frac{a_n}{(2n-1)!} p^{2n-1}
\label{eq:borel0}
\end{eqnarray}
The formal  $t\to+\infty$ series for $h(t)$ in (\ref{eq:h}) is recovered by the Laplace transform:
\begin{eqnarray}
\text{inverse Borel transform:} \qquad h(t)\sim \int_0^\infty dp\, e^{-p\, t} \mathcal B [h](p)
\label{eq:inverse-borel0}
\end{eqnarray}
Thus, the task of extrapolating and analytically continuing the function $h(t)$ [and therefore also the Painlev\'e solution $y(x)$] along the real axis, and  into the complex plane, becomes the problem of understanding the singularity structure and analytic continuation of the Borel transform $\mathcal B [h](p)$ in the complex $p$-plane, the Borel plane. The pragmatic question is: 
\begin{quote}
How much can we learn about the Borel function $\mathcal B [h](p)$ from just a {\it finite} number of perturbative coefficients of the small $p$ expansion  defined in (\ref{eq:borel0})?
\end{quote}

\subsection{Pad\'e Analysis of the Borel Transform}
\label{sec:pade-borel}

\begin{figure}[htb]
\centering{\includegraphics[scale=0.7]{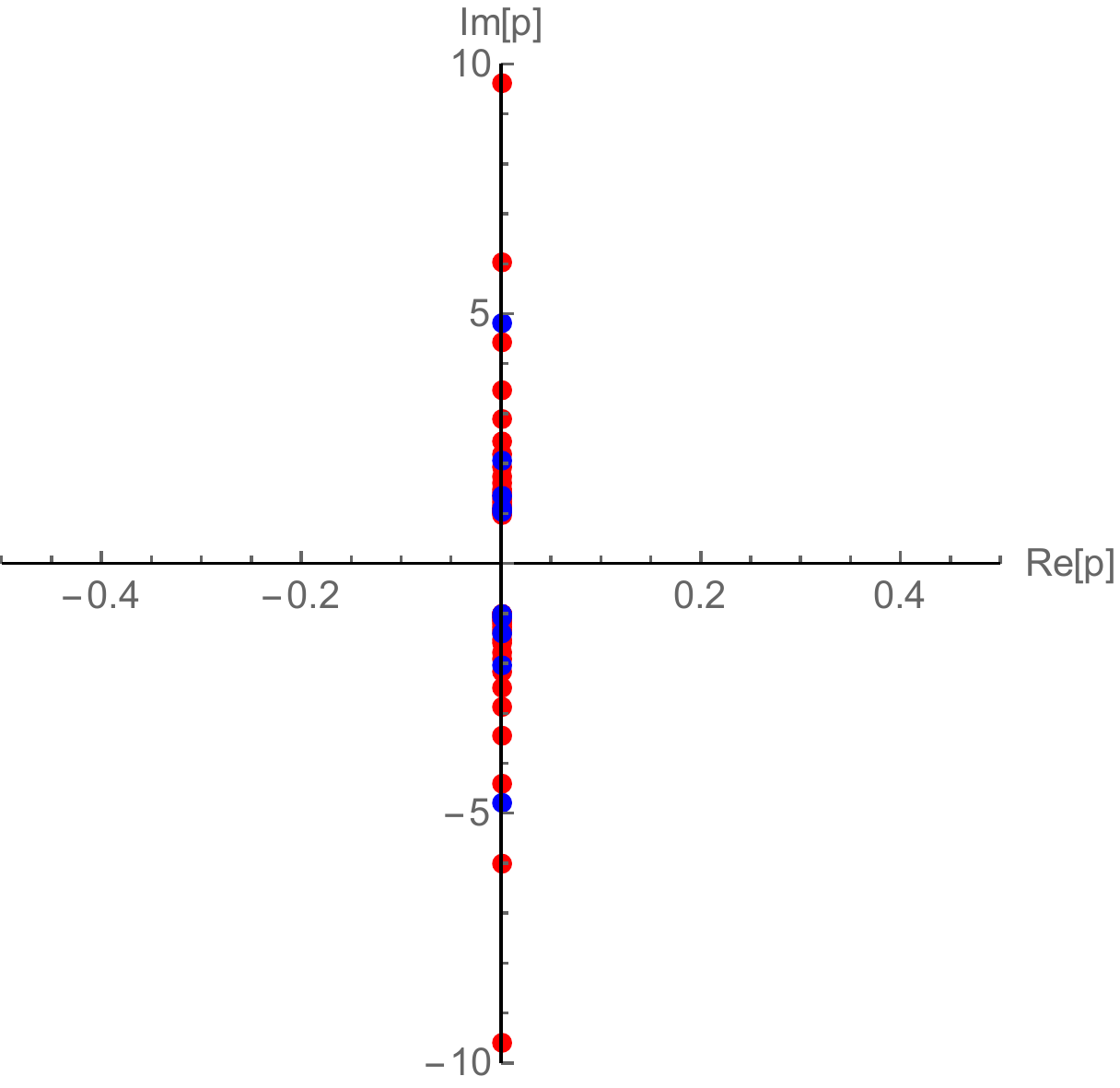}}
\caption{Poles of the Pad\'e approximation ${\mathcal {PB}}_{N}[h](p)$ in (\ref{eq:pade-borel0}) for the truncated Borel transform. The blue points are the Pad\'e poles with $N=10$ input coefficients, and the red points represent the Pad\'e poles with $N=50$ input coefficients. (Some further poles of ${\mathcal {PB}}_{50}[h](p)$ are not shown, due to scale). Note that all the poles lie on the imaginary axis, and accumulate to branch points at $p=\pm i$.}
\label{fig:pade-borel-poles}
\end{figure}

The Borel transform series (\ref{eq:borel0}) has a finite radius of convergence, equal to 1,  in the Borel $p$ plane. This follows from the large-order growth in (\ref{eq:factorial}), and can also be seen from a simple ratio test \cite{fisher-series}. It is also encoded  in the distribution of poles of a Pad\'e approximation to the Borel transform: see Fig. \ref{fig:pade-borel-poles}, which indicates singularities at $p=\pm i$. Given $N$ coefficients in a truncated Borel transform
\begin{eqnarray}
\text{truncated Borel transform:} \qquad \mathcal B_{N} [h](p)\equiv \sum_{n=1}^{N} \frac{a_n}{(2n-1)!}\, p^{2n-1}
\label{eq:borel1}
\end{eqnarray}
and noting that it is  a polynomial of degree $(2N-1)$,
we form the off-diagonal $[N-1, N]$ Pad\'e approximant \cite{graffi-simon,baker,pade,carl-book}
\begin{eqnarray}
\text{Pad\'e-Borel transform:} \qquad \mathcal{PB}_{N} [h](p) = \frac{P_{N-1}(p)}{Q_N(p)}
\label{eq:pade-borel0}
\end{eqnarray}
where $P_{N-1}(p)$ and $Q_N(p)$ are polynomials of degree $(N-1)$ and $N$, respectively.
This Pad\'e approximant step is completely algorithmic, and is indeed a built-in function in Mathematica and Maple. The Pad\'e-Borel poles are shown in Fig. \ref{fig:pade-borel-poles}, for $N=10$, and for $N=50$. These poles are interlaced by the associated Pad\'e zeros. Pad\'e approximants represent a branch cut by a string of interlaced poles and zeros, so the Pad\'e-Borel transform suggests the existence of branch cuts along the imaginary axes, with branch points at $p=\pm i$.  

As a side-comment, we note that we have found that it is more numerically stable, especially when dealing with larger values of $N$, to convert the Pad\'e approximant to its partial fraction decomposition: 
\begin{eqnarray}
\text{Partial fraction Pad\'e-Borel transform:} \quad \mathcal{PB}_{N} [h](p) = \frac{P_{N-1}(p)}{Q_N(p)}
=\sum_{k=1}^{N} \frac{r_k}{p-p_k}
\label{eq:pf-pade-borel1}
\end{eqnarray}
where the sum is over the $N$ Pad\'e poles $p_k$, the zeros of $Q_N(p)$, with associated residues $r_k$.
 \begin{figure}[htb]
\centering{\includegraphics[scale=.65]{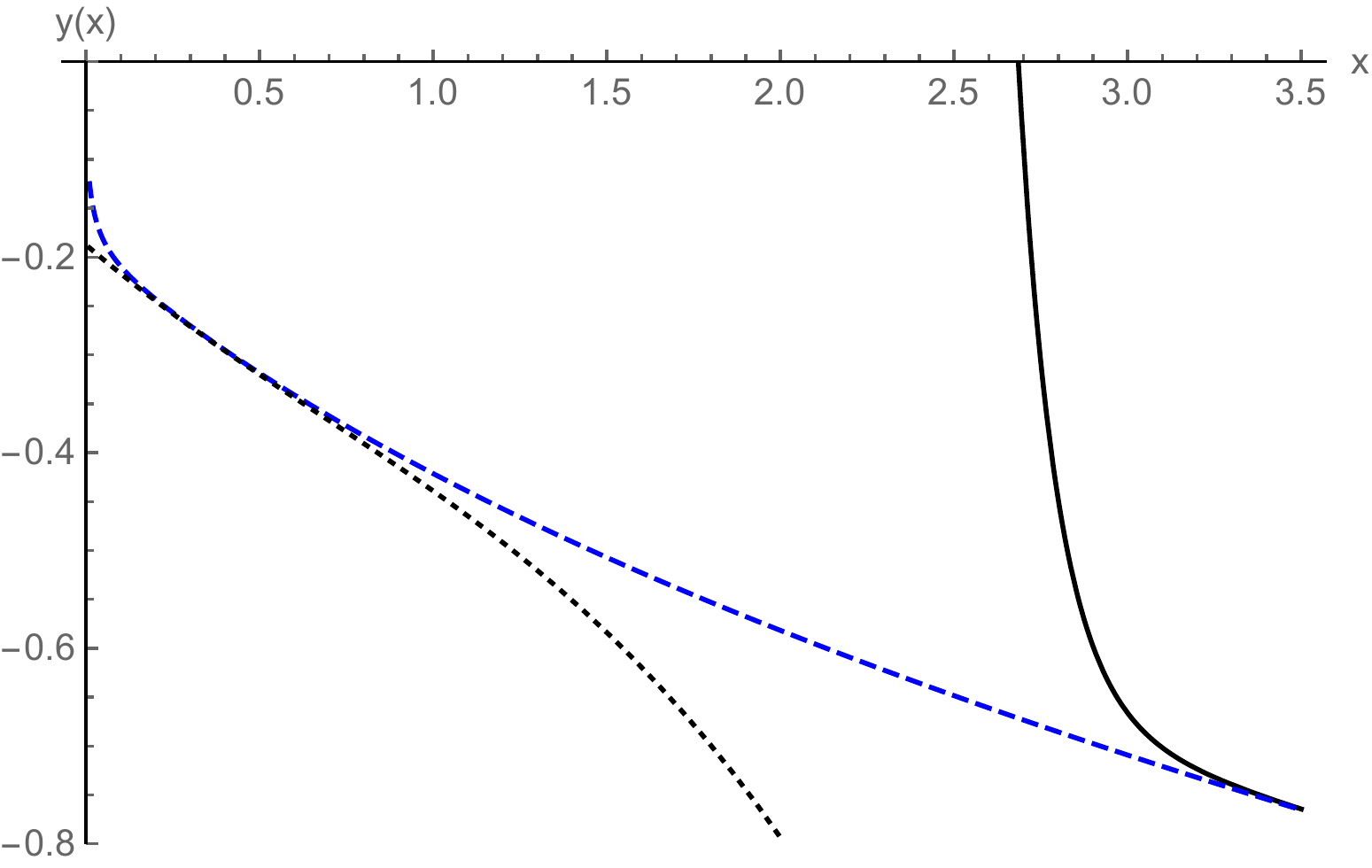}}
\caption{A plot of the extrapolation of the PI function $y(x)$, from $x=+\infty$ down to near $x=0$, starting with just $N=10$ terms in (\ref{eq:input}) of the asymptotic expansion at $x=+\infty$. The solid black curve is based on (\ref{eq:inverse-borel0}) with the truncated Borel transform form $\mathcal B_{10} [h](p)$ defined in (\ref{eq:borel1}), while the dashed blue curve is based on (\ref{eq:inverse-borel0}) with the 
Pad\'e approximant ${\mathcal PB}_{10} [h](p)$ defined in (\ref{eq:pade-borel0}). The dotted black curve shows the first few terms of the Taylor expansion at $x=0$, using the initial conditions for the PI {\it tritronqu\'ee} solution from \cite{joshi}: $y(x)\approx -0.1875543083 - 0.3049055603 x + 0.1055298557 x^2 - 0.05229396374 x^3$. Note that the Pad\'e-Borel transform extrapolates $y(x)$ to much smaller values of $x$ than does the 'raw' Borel transform. Starting with more input coefficients (i.e. larger values of $N$), the improvement gets even better.
}
\label{fig:y-pade-borel-10}
\end{figure}
Since the Pad\'e approximant is a rational function of $p$,  the partial fraction expression in (\ref{eq:pf-pade-borel1}) is in principle equivalent to the Pad\'e expression in (\ref{eq:pade-borel0}), but the increased numerical stability of (\ref{eq:pf-pade-borel1}) arises because the Pad\'e expression at large order $N$ tends to have very large coefficients in both numerator and denominator, thereby causing instabilities due to massive cancellations. In contrast,  the residues and poles in the partial fraction expression (\ref{eq:pf-pade-borel1}) are  much smaller in magnitude, and so the evaluation is more stable. This is a typical instability of Pad\'e approximants. The resulting improved numerical stability  of the partial fraction form can be important for subsequent numerical  integrations over $p$, such as the Laplace transform integral in (\ref{eq:inverse-borel0}), required for returning to the original functions $h(t)$ and $y(x)$ in the physical $t$ and $x$ planes.

The Pad\'e-Borel approximant in (\ref{eq:pade-borel0})-(\ref{eq:pf-pade-borel1}) is simple to compute, and it provides a significant extrapolation  down  the positive real line of the asymptotic expansion (\ref{eq:largex}) of the PI solution $y(x)$. This is illustrated in Fig. \ref{fig:y-pade-borel-10}, in which we plot the extrapolation obtained starting with $N=10$ terms.  As a diagnostic comparison, we also plot [black dotted curve] the first four terms of the Taylor expansion of  $y(x)$ at the origin, using the initial conditions at the origin for the PI {\it tritronqu\'ee} solution \cite{joshi}: $y(x)\approx -0.1875543083 - 0.3049055603 x + 0.1055298557 x^2 - 0.05229396374 x^3+\dots$.  Fig. \ref{fig:y-pade-borel-10} strongly suggests that the Pad\'e-Borel extrapolation approaches the {\it tritronqu\'ee} solution as $x\to 0$, while the "raw" Borel extrapolation begins to diverge from this behavior at $x\approx 3$. 
So, just 10 terms at $x\to+\infty$ are required for the Pad\'e-Borel extrapolation to extrapolate accurately down to $x\approx 0.25$.
In the next Section we show that by combining the Pad\'e-Borel method with conformal mapping we obtain a dramatic improvement on this already impressive precision of the Pad\'e-Borel extrapolation.

\begin{figure}[htb]
\centering{\includegraphics[scale=.4]{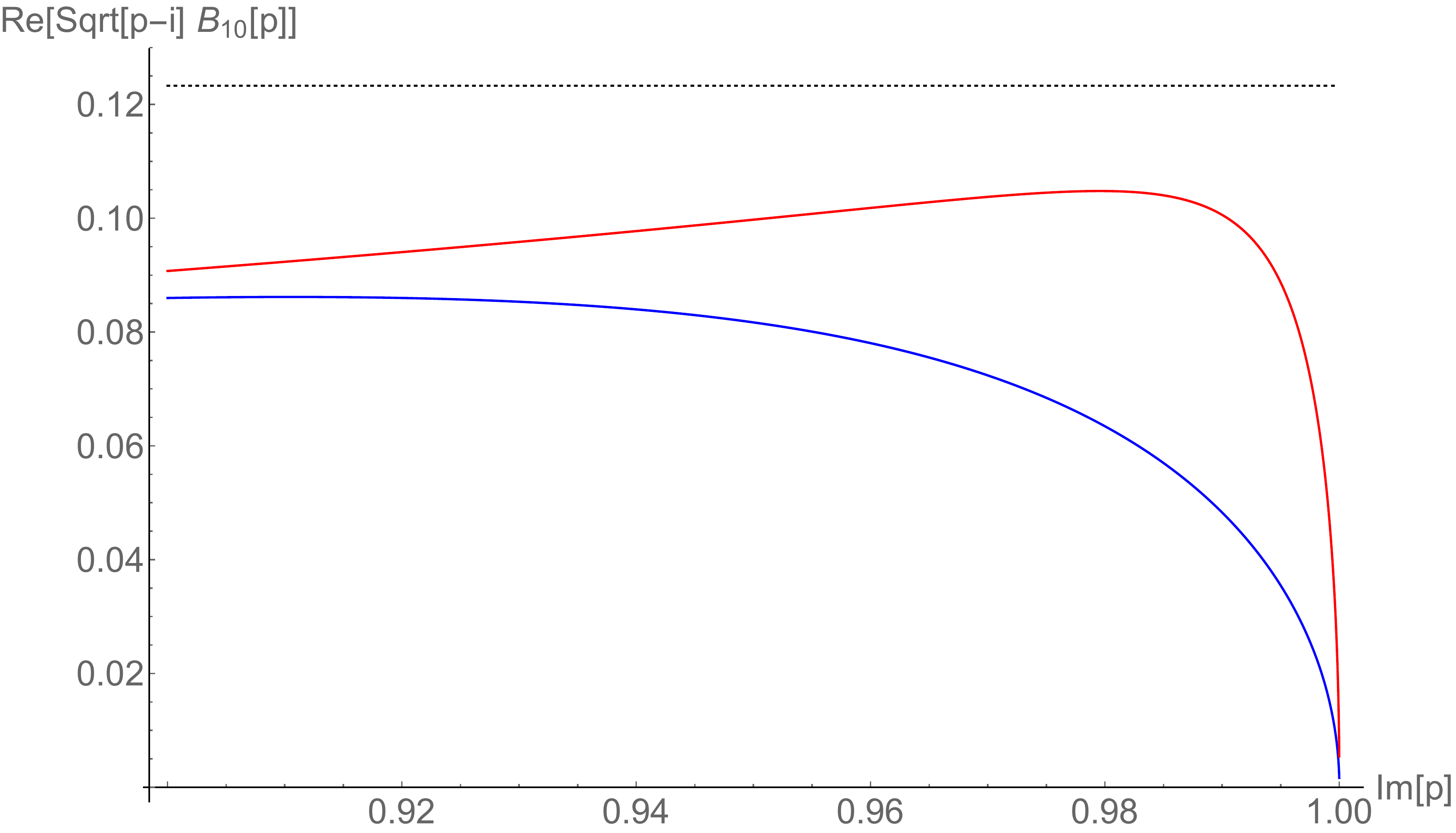}}
\caption{An illustration of the square root branch cut behavior (\ref{eq:nature}) of the Borel function as the leading singularity at $p=+i$ is approached. The  plot shows $\text{Re}[\sqrt{(p-i)}\,\mathcal{B}_{10}[h](p)]$ based on the $N=10$ Borel transform (\ref{eq:borel1}) [blue curve], and  based on the corresponding Pad\'e-Borel transform (\ref{eq:pade-borel0}) [red curve]. For reference, the dotted horizontal line is  half the value of the Stokes constant, $\frac{1}{2\pi}\sqrt{\frac{3}{5}}$. This approach to the leading singularity is probed more precisely below, after the use of a suitable conformal map: see Figure \ref{fig:conformal-pade-borel-stokes-10}.}
\label{fig:pade-borel-stokes-10}
\end{figure}

Having determined the {\it location} of the leading singularities, the next step is to determine the {\it nature} of these singularities. There are several complementary ways to do this. A simple method is to use Darboux's theorem, which relates the nature of singularities to the large-order growth of coefficients of expansions about some other point \cite{henrici,guttmann}. The large-order growth in (\ref{eq:factorial}) implies that the leading singularities at $p=\pm i$ are square root singularities: 
\begin{eqnarray}
  {\mathcal B}[h](p)\sim \frac{c}{\sqrt{p\mp i}}
   \qquad, \qquad p\to \pm i 
\label{eq:nature}
\end{eqnarray}
for some constant $c$. 
This can also be extracted from  the distribution of the interlacing Pad\'e poles and zeros, but the Darboux analysis is simpler.
Furthermore, invoking resurgence, the constant $c$ in (\ref{eq:nature}) is directly related to the Painlev\'e I Stokes constant: $c=\frac{1}{2}S=\frac{1}{2\pi} \sqrt{\frac{3}{5}}$. In Fig. \ref{fig:pade-borel-stokes-10} 
we plot the approach to the leading singularities, at $p=\pm i$, of the Pad\'e-Borel approximation (\ref{eq:pade-borel0}) to the Borel function, for $N=10$.
Fig. \ref{fig:pade-borel-stokes-10}  displays rough numerical evidence for the approach to half the known Stokes constant for PI, marked as a horizontal dotted line. The Pad\'e-Borel transform (\ref{eq:pade-borel0}) is much better near the singularity than the raw Borel transform (\ref{eq:borel1}), but the behavior very close to the singularity still deviates from this resurgent expectation. With more input coefficients (larger $N$) the curves approach closer to the Stokes value, but the deviation persists close to the singularity. This will be probed to significantly higher precision in the next Section, in which we introduce more powerful tools than just Pad\'e-Borel. Compare, for example, with Fig. \ref{fig:conformal-pade-borel-stokes-10}.

\subsection{Conformal Mapping of the Borel Plane}
\label{sec:conformal}

\begin{figure}[htb]
\centering{\includegraphics[scale=.5]{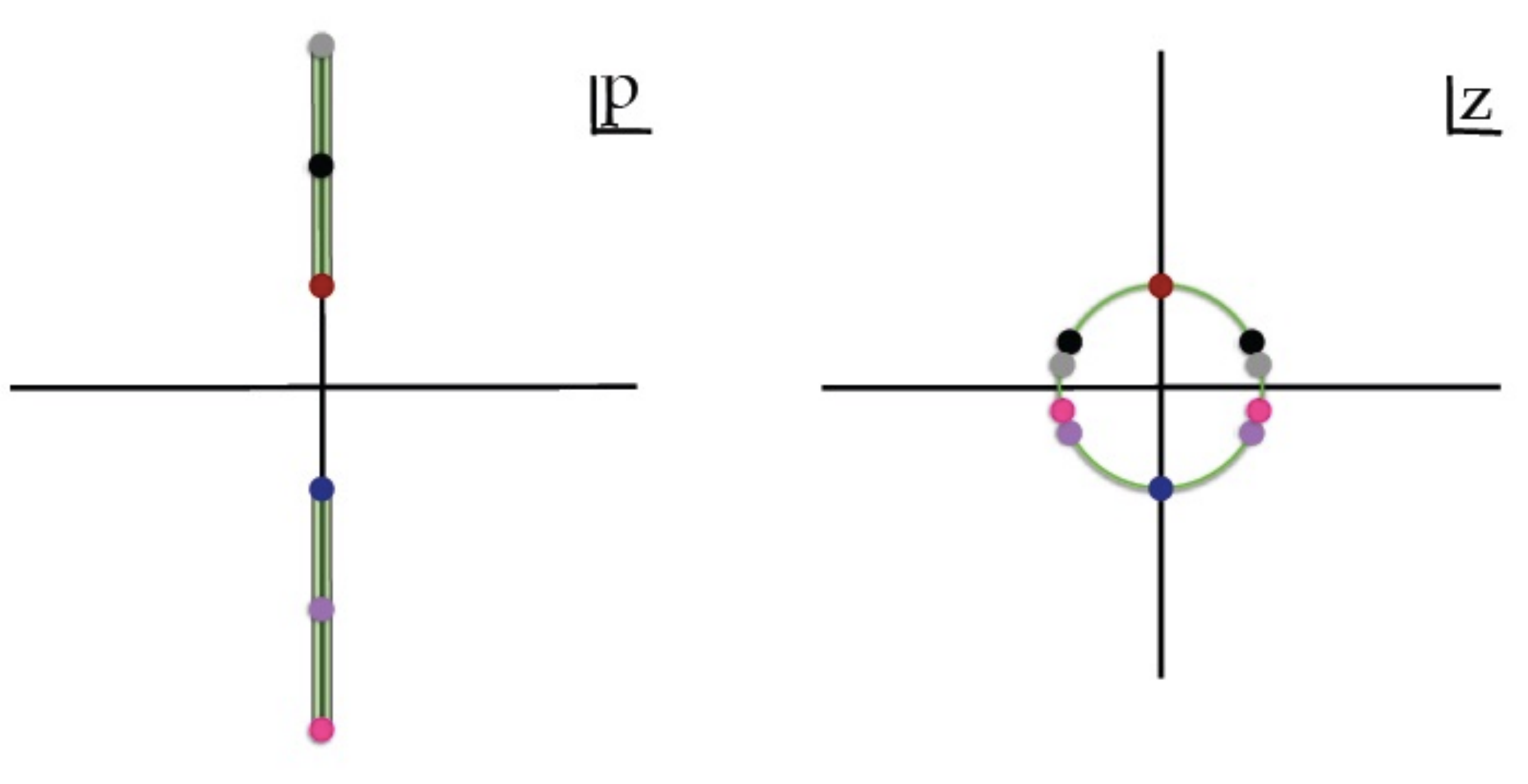}}
\caption{The conformal map (\ref{eq:map1}) from the doubly-cut Borel $p$ plane [left] to the unit disc in the $z$-plane [right]. The cuts on the imaginary axis in the $p$ plane are mapped to the unit circle in the $z$ plane, marked by the green lines. The leading branch points at $p=\pm i$ map to $z=\pm i$, marked by the red and blue dots. The resurgent second singularities at $p=\pm 2i$, on either side of the cuts, map to $z=\pm e^{\pm i\pi/6}$, marked by the black and purple  dots, and the third singularities at $p=\pm 3i$ map to $\pm\frac{1}{3}(\sqrt{8}\pm i)$, marked by the grey and pink dots.}
\label{fig:conformal-map}
\end{figure}
To refine our numerical exploration of the properties of the Borel transform function $\mathcal{B}[h](p)$, we define a conformal map from the doubly-cut Borel $p$ plane into the unit disc in the $z$ plane:
\begin{eqnarray}
\text{conformal map:}\qquad z=\frac{p}{1+\sqrt{1+p^2}}\qquad \longleftrightarrow \qquad p=\frac{2\, z}{1-z^2}
\label{eq:map1}
\end{eqnarray}
See Fig. \ref{fig:conformal-map}. This maps the cuts along the imaginary axis in the Borel $p$ plane to the unit circle in the complex $z$ plane. The leading singularity at $p=\pm i$ is mapped to $z=\pm i$. The resurgent second singularity at $p=\pm 2i$, on either side of the cut, is mapped to $z=\pm e^{\pm i\pi/6}$.
Similarly, the resurgently repeated singularities at $p_k\equiv \pm k\, i$, map to points on the unit circle at $z_k\equiv \pm \frac{1}{k}(\sqrt{k^2-1}\pm i\, k)$, which approach $z=\pm 1$ as $k\to\infty$. Thus the point at infinity in the $p$ plane maps to $z=\pm 1$.  Conformal mapping combined with Borel transforms is a well-known technique in physical applications \cite{kleinert,caliceti,thooft,kazakov,zinn-qft}, and here we quantify its improvement over the Pad\'e-Borel transform discussed in the previous section, and we also use resurgent asymptotics to explain why such a significant improvement occurs.

We define a Pad\'e-Conformal-Borel transform by the following algorithmic  steps: 
\begin{enumerate}
\item
For a given number $N$ of input coefficients, we evaluate the truncated Borel transform  $\mathcal{B}_{N} [h](p)$ in (\ref{eq:borel1}) at the conformally mapped location, $p=\frac{2\, z}{1-z^2}$: 
\begin{eqnarray}
\text{Conformal-Borel transform:} \qquad \mathcal{CB}_{N} [h](z)\equiv \mathcal{B}_{N} [h]\left(\frac{2\, z}{1-z^2}\right)
 \label{eq:conformal-borel0}
\end{eqnarray}
By construction, this function is analytic inside the unit disc in the $z$-plane. 
\item
Re-expand $\mathcal{CB}_{N} [h](z)$ about $z=0$ to the same order as the original Borel transform $\mathcal{B}_{N} [h](p)$ in $p$ [i.e., to $O(z^{2N-1})$], and then construct a  Pad\'e approximant of the resulting truncated Taylor expansion, in the $z$ plane:
\begin{eqnarray}
&~&\text{$z$-plane Pad\'e-Conformal-Borel transform:}\nonumber\\
&&\quad  \mathcal{PCB}_{N}[h](z)\equiv \text{Pad\'e}_N\left\{ \text{Taylor expansion of}\, \mathcal{CB}_{N} [h](z)\right\} 
\label{eq:pade-conformal-borel-z}
\end{eqnarray}
\item
Invert the conformal map, evaluating the $z$-plane Pad\'e approximant (\ref{eq:pade-conformal-borel-z}) at $z=\frac{p}{1+\sqrt{1+p^2}}$:
\begin{eqnarray}
&~&\text{$p$-plane Pad\'e-Conformal-Borel transform:}
\nonumber\\
&&\quad  
\mathcal{PCB}_{N}[h](p)\equiv \mathcal{PCB}_{N} [h]\left(z=\frac{p}{1+\sqrt{1+p^2}}\right) 
\label{eq:pade-conformal-borel-p}
\end{eqnarray}
\end{enumerate}
These steps of conformal mapping, followed by re-expansion and Pad\'e approximation, and inverting the conformal map, yield our global approximation (\ref{eq:pade-conformal-borel-p}) to the Borel transform function in the original Borel $p$ plane, which we call the ``Pad\'e-Conformal-Borel transform''. We stress that these steps are purely algorithmic, and furthermore they do not rely on any special integrability properties of the Painlev\'e I equation. In the next Sections we illustrate the numerical  advantages of  this conformal mapping procedure.

\subsection{Increased Precision from the Pad\'e-Conformal-Borel Transform}
\label{sec:pcb-improvement}

\begin{figure}[htb]
\centering{\includegraphics[scale=.8]{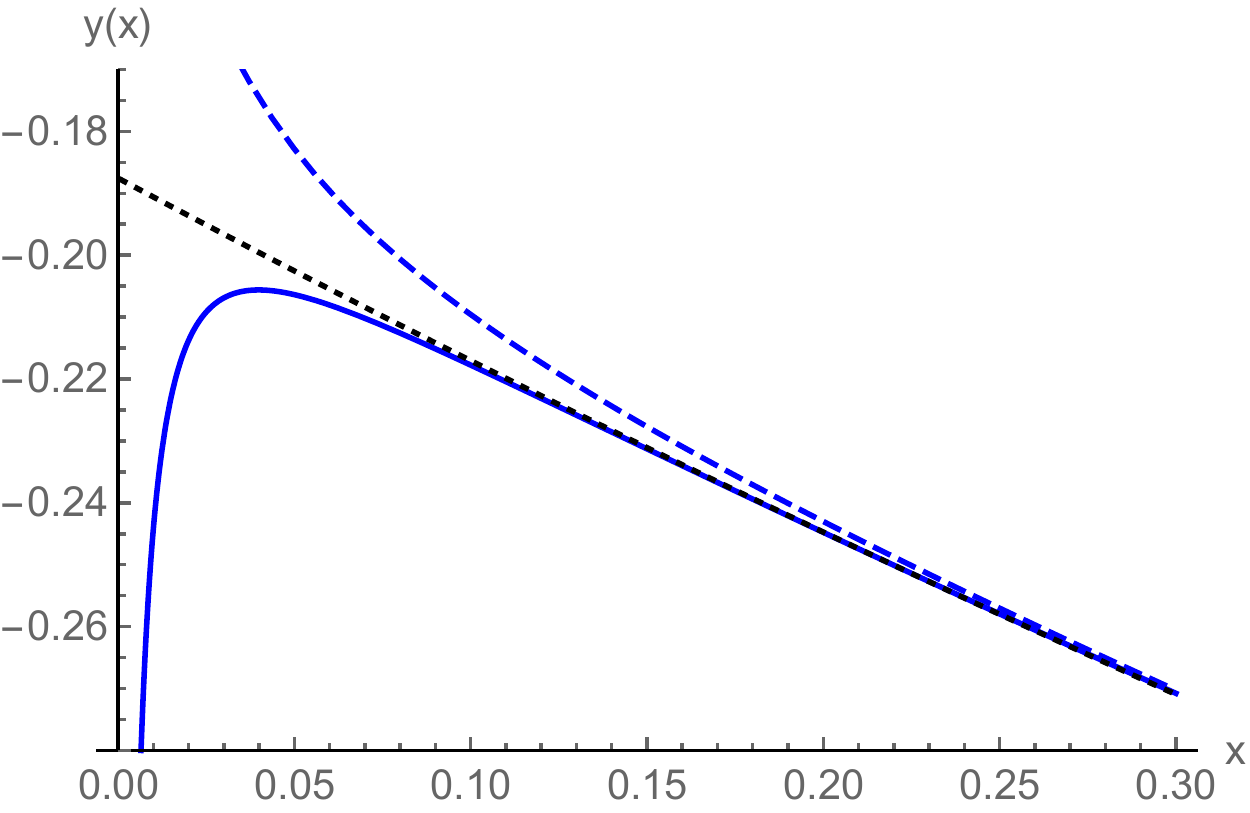}}
\caption{The solid blue curve shows the PI function $y(x)$ reconstructed from the Pad\'e-Conformal-Borel transform $\mathcal{PCB}_{N}[h](p)$ defined in (\ref{eq:pade-conformal-borel-p}), with $N=10$ input coefficients. For comparison, the dashed blue curve shows the result for  $y(x)$ from using the Pad\'e-Borel transform $\mathcal{PB}_{10} [h](p)$, without the conformal mapping, as shown in Fig. \ref{fig:y-pade-borel-10}. On this scale the original 10-term truncated Borel transform (\ref{eq:borel1}) (the solid black curve in Fig.  \ref{fig:y-pade-borel-10}) does not even appear.  And as in Fig. \ref{fig:y-pade-borel-10}, the dotted black curve shows the first few terms of the Taylor expansion at the origin, using the 
{\it tritronqu\'ee} initial conditions. Note that both Borel transforms provide a remarkably good extrapolation from 
$x=+\infty$ down to $x\approx .25$, with only 10 input coefficients in (\ref{eq:input}).
However, with the same amount of input data, we see that the conformal-Pad\'e-Borel extrapolation [solid blue curve] is significantly better at even smaller values of $x$ than the Pad\'e-Borel extrapolation [dashed blue curve]. See Fig. \ref{fig:y-conformal-pade-borel-50} for the analogous plot with $N=50$ input coefficients.}
\label{fig:y-conformal-pade-borel-10}
\end{figure}
\begin{figure}[htb]
\centering{\includegraphics[scale=.8]{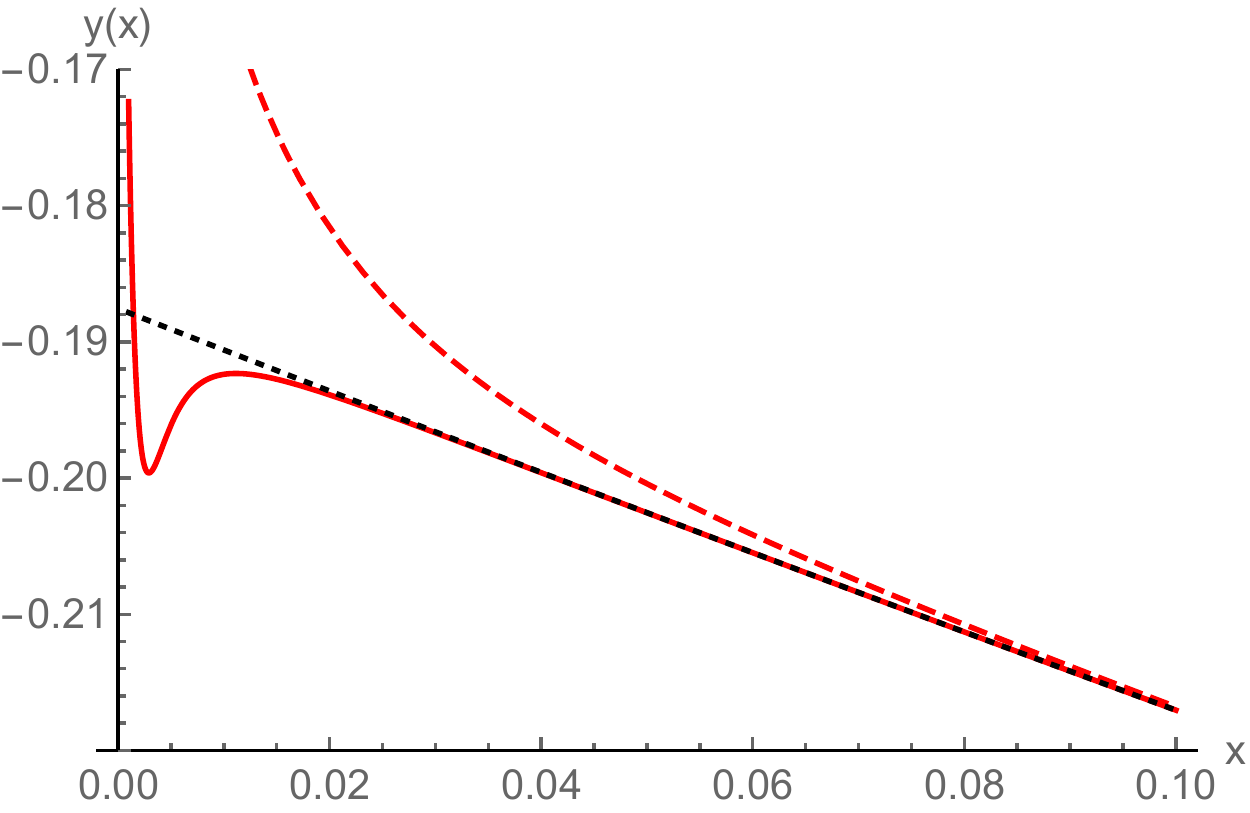}}
\caption{As in Fig. \ref{fig:y-conformal-pade-borel-10}, but now starting with  $N=50$ perturbative input coefficients. The solid red curve shows the Pad\'e-Conformal-Borel extrapolation of the PI function $y(x)$, starting from $N=50$ terms of the asymptotic large $x$ expansion in (\ref{eq:largex}), while the dashed red curve shows the extrapolation based on the Pad\'e-Borel transform, without the conformal mapping. Once again, the conformal mapping extends the extrapolation to even smaller values of $x$. The black dotted curve shows the first few terms of the Taylor expansion of the PI {\it tritronqu\'ee} solution at the origin.}
\label{fig:y-conformal-pade-borel-50}
\end{figure}

The first indication of higher precision using the Pad\'e-Conformal-Borel transform, compared to the 
Pad\'e-Borel transform, comes from performing the numerical inverse Borel transform integration in (\ref{eq:inverse-borel0}) to reconstruct the function $h(t)$, and hence the Painlev\'e I solution $y(x)$ in the original $x$ plane using (\ref{eq:ecallet})-(\ref{eq:yh}). This leads to the numerical evaluation of an extrapolation of the PI solution $y(x)$ along the positive $x$ axis, using as input $N$ terms of its asymptotic expansion (\ref{eq:largex}) at $x\to +\infty$:
\begin{eqnarray}
y_N(x)=-\sqrt{\frac{x}{6}}\left(1+\int_0^\infty dp\, \mathcal{B}_{N}[h](p)\, \exp\left[-p\,\frac{\left(24\, x\right)^{5/4}}{30} \right]\right)
\label{eq:num-y}
\end{eqnarray}
Here we can replace the Borel transform function $\mathcal{B}_{N}[h](p)$ by either its Pad\'e-Borel $\mathcal{PB}_{N}[h](p)$ or Pad\'e-Conformal-Borel $\mathcal{PCB}_{N}[h](p)$ approximation, defined in (\ref{eq:pade-borel0}) or (\ref{eq:pade-conformal-borel-p}) respectively.
 Figs. \ref{fig:y-conformal-pade-borel-10} and \ref{fig:y-conformal-pade-borel-50} compare the reconstructed function $y_N(x)$ using the Pad\'e-Borel transform [dashed curves], and using the 
 Pad\'e-Conformal-Borel transform [solid curves], for $N=10$ and $N=50$, respectively. For reference, the dotted black line shows  the (convergent) expansion of $y(x)$ at the origin, using the approximate {\it tritronqu\'ee} initial values \cite{joshi}: $y(0)\approx  -0.1875543...$ and $
y'(0)\approx  -0.34090556 ....$. 
 Figs. \ref{fig:y-conformal-pade-borel-10} and \ref{fig:y-conformal-pade-borel-50} show that {\it both} Borel transforms, $\mathcal{PB}_{N}[h](p)$ and $\mathcal{PCB}_{N}[h](p)$, lead to remarkably precise extrapolations from $x=+\infty$ down to very small values of $x$, even though the extrapolations are based on just 10 or 50 input coefficients from the asymptotic expansion at $x=+\infty$. However, we observe that in both cases the Pad\'e-Conformal-Borel transform leads to a higher precision extrapolation at small $x$.
 In Section \ref{sec:re-expand} we show how to obtain even better precision all the way to $x=0$, and also into the negative $x$ region and the complex $x$ plane.

\section{Extrapolation into the Complex Plane: Resurgence}
\label{sec:complex}

In this Section we discuss how and why the Pad\'e-Conformal-Borel transform $\mathcal{PCB}_{N} [h](p)$ in (\ref{eq:pade-conformal-borel-p}) provides a much more precise representation of the true Borel transform function than the Pad\'e-Borel transform $\mathcal{PB}_{N} [h](p)$ in (\ref{eq:pade-borel0}). The source of this improvement is the resurgent structure underlying the original asymptotic expansion, which is encoded more precisely by the Pad\'e-Conformal-Borel transform. While the two Borel functions  $\mathcal{PCB}_{N} [h](p)$ and $\mathcal{PB}_{N} [h](p)$ are very similar along the positive real $p$ axis, important differences arise in the complex $p$ plane, especially as one approaches  the singularity lines on the imaginary $p$ axis (see, for example, Figs. \ref{fig:borel-two-jumps-p50},  \ref{fig:pade-borel-no-jump}, and \ref{fig:jump-50}, below). This affects the precision with which the contour of the Laplace transform in (\ref{eq:num-y}) can be deformed, thereby restricting the region of the complex $x$ plane into which  the Painlev\'e I function $y(x)$ can be analytically continued with precision.

\subsection{Precision Evaluation of the Painlev\'e I Stokes Constant}
\label{sec:stokes}
\begin{figure}[htb]
\centering{\includegraphics[scale=.4]{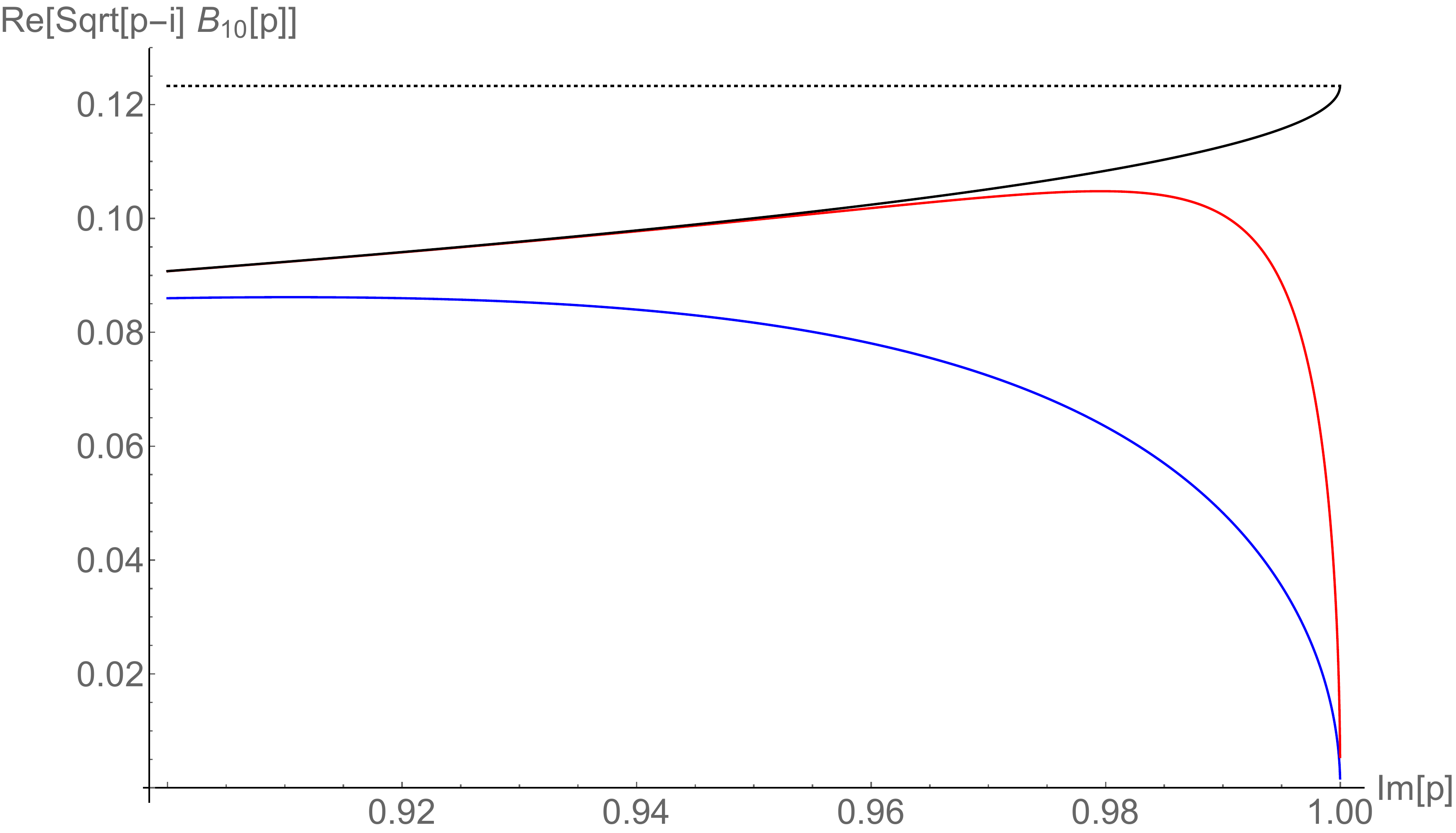}}
\caption{An illustration of the square root branch cut behavior (\ref{eq:nature}) of the Borel function as the leading singularity at $p=+i$ is approached. The  plot shows $\text{Re}[\sqrt{(p-i)}\,\mathcal{B}_{10}[h](p)]$ based on the $N=10$ Borel transform [blue curve], Pad\'e-Borel transform [red curve] and Pad\'e-Conformal-Borel transform [black curve]. For reference the dotted horizontal line is  half the value of the Stokes constant, $\frac{1}{2\pi}\sqrt{\frac{3}{5}}$. }
\label{fig:conformal-pade-borel-stokes-10}
\end{figure}
Invoking resurgence of the PI function $y(x)$, the behavior of the Borel transform function near its first singularities at $p=\pm i$ determines the Painlev\'e I Stokes constant, and therefore governs the associated exponential corrections. This can be used as a precise numerical test of resurgence. Recall from Fig. \ref{fig:pade-borel-stokes-10} that the Pad\'e-Borel function $\mathcal{PB}_{N} [h](p)$ shows hints of the expected square root singularity behavior near the first singularity, but deviates as $p$ approaches very close to $\pm i$. After conformal mapping, the Pad\'e-Conformal-Borel transform is dramatically more precise in the vicinity of these first singularities, even with just $N=10$ input coefficients, as shown in Fig. \ref{fig:conformal-pade-borel-stokes-10}. In fact, starting with just $N=10$ coefficients, the behavior of the Pad\'e-Conformal-Borel transform near the leading singularity determines the Stokes constant to 4 digits of precision. And with $N=50$ digits we obtain 23 digits of precision. By contrast, probing the Pad\'e-Borel transform near its first singularities allows just 1  digit of precision, at best. This is a dramatic improvement, which translates into a significant enlargement of the region of the complex plane in the physical variable $x$ that can be accurately probed using the inverse Borel transform (\ref{eq:num-y}).

\subsection{Resurgent Structure of Poles of the Pad\'e-Conformal-Borel Transform}
\label{sec:res-poles}
\begin{figure}[htb]
\centering{\includegraphics[scale=.7]{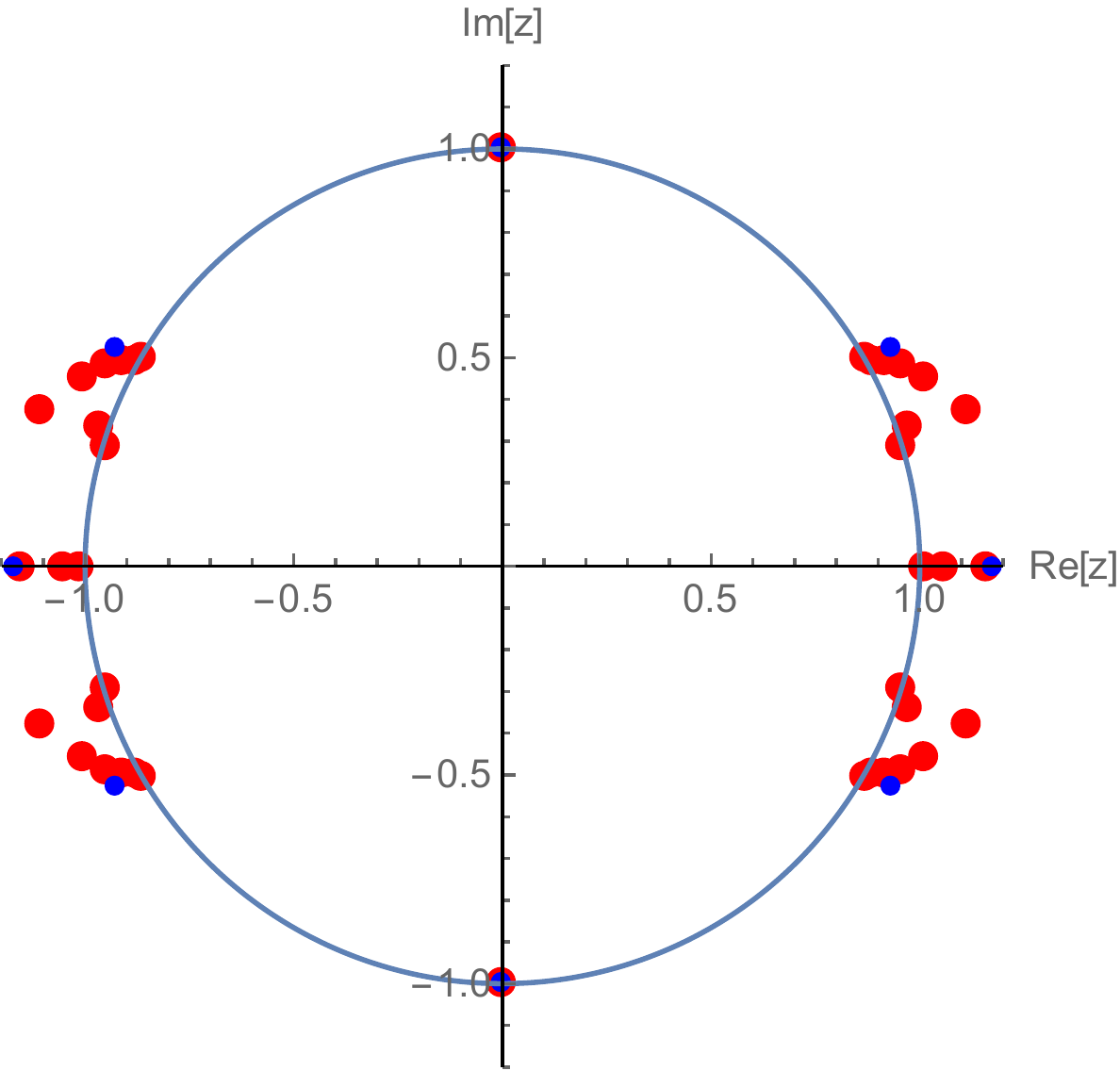}}
\caption{Poles in the $z$ plane of the conformally mapped Borel transform, for $N=10$ terms [blue] and 50 terms [red].}
\label{fig:z10}
\end{figure}
The underlying reason for the remarkable improvement of the Pad\'e-Conformal-Borel transform is that the  conformal map reveals the resurgent structure in the Borel plane, which is hidden in the pole structure of the Pad\'e-Borel transform shown in Fig. \ref{fig:pade-borel-poles}. The conformal map separates and resolves the sub-structure of the Borel singularities, showing the repetition of singularities at integer multiples of the first singularities, as expected for a resurgent solution to a non-linear differential equation such as the  Painlev\'e I equation (\ref{eq:p1}) \cite{ecalle,costin-dmj,costin-book,costin-odes,costin-costin-inventiones,Aniceto:2013fka,Dorigoni:2014hea,sauzin,gokce}.

To see this, consider  the poles in the conformally mapped $z$ plane of the Pad\'e-Conformal-Borel function $\mathcal{PCB}_{N}[h](z)$ defined in (\ref{eq:pade-conformal-borel-z}). These are shown in Fig. \ref{fig:z10}, for $N=10$ and $N=50$. These poles display the resurgent structure of the problem. The poles at $z=\pm i$ are the conformal map images of the leading square root branch points at $p=\pm i$ in the Borel $p$ plane.
Significantly, the conformal map has converted these branch cut singularities to simple poles, with residue proportional to the PI Stokes constant. Furthermore, these leading poles have been separated from the other singularities. This separation of the singularities according to their resurgent structure is a key factor in the improved precision of the subsequent numerical evaluations. Further observations concerning the $z$ plane poles of the Pad\'e-Conformal-Borel function $\mathcal{PCB}_{N}[h](z)$ are:
\begin{enumerate}
\item
There are no poles inside the unit disc in the $z$ plane, since for any $N$ the function 
$\mathcal{PCB}_{N}[h](z)$ is analytic inside the unit disc, by construction\footnote{In general, 
 Pad\'e may produce spurious poles, for example with anomalously small residues; these can be filtered if necessary \cite{trefethen}, but we did not encounter this situation in this problem.}. The poles appear on or outside the boundary of the unit disc. 
\item
The poles on the {\it boundary} of the unit disc in the $z$ plane are the conformal map images of  $p$ plane singularities on the cuts  along the imaginary $p$ axis. 
\item
The leading square-root singularities at $p=\pm i$ have been  conformally mapped to simple poles at $z=\pm i$, whose residue determines the Stokes constant $S=\frac{1}{\pi}\sqrt{\frac{3}{5}}$ with remarkable precision. (Note that if the leading singularity were not of square root form, the conformal map would not map it to a pole; however, mapping to a pole can be achieved by suitable re-definition and convolution transformations. See the discussion in the Conclusions.)
\item
The poles accumulating at $z=\pm e^{\pm i\,\pi/6}$ are the conformal map images of singularities accumulating (from either side of the cuts) at $p=\pm 2i$ in the $p$ plane. We also see an indication of poles accumulating at $z=\pm  \frac{1}{3}(\sqrt{8} \pm i)$, which are the conformal map images of singularities at $p=\pm 3i$ in the $p$ plane. With higher values of $N$, further $p$ plane singularities at higher integer multiples of $\pm i$ are resolved.
\item
The poles outside the unit disc in the $z$ plane correspond to information about the Borel transform on higher Riemann sheets, arising from analytic continuation across the $p$ plane  cuts. This information can be used for further numerical refinement.
\end{enumerate}
\begin{figure}[htb]
\centering{\includegraphics[scale=.7]{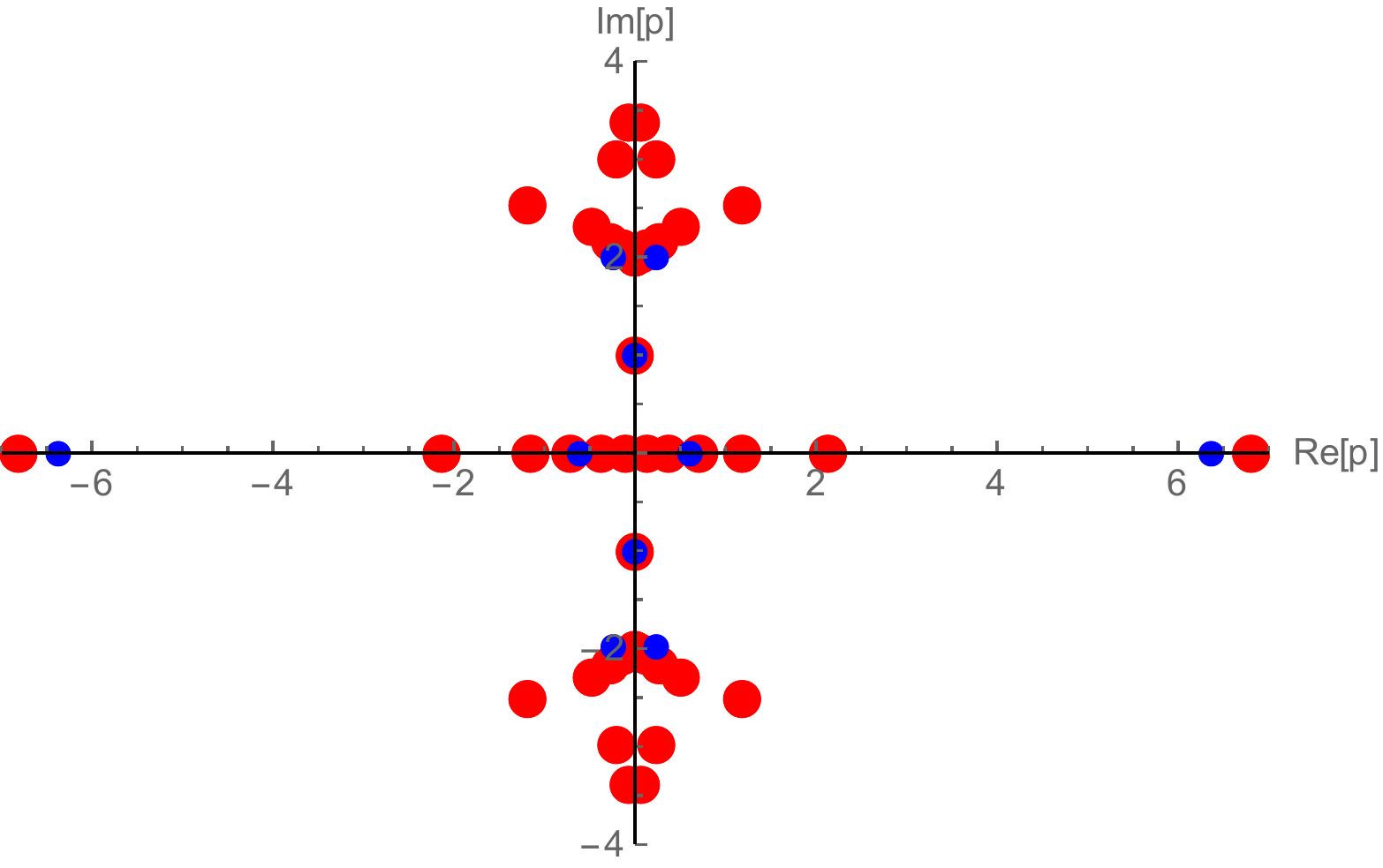}}
\caption{The $z$-plane poles mapped back to the complex Borel $p$ plane: 10 term expansion [blue], and 50 term expansion [red]. The leading singularity at $p=\pm i$ is separated from the other singularities, and the next singularities accumulate at $p=\pm 2i$, the resurgently repeated next singularity, and with $N=50$ we see a hint of accumulation at the next singularity $p=\pm 3i$. The singularities on the real $p$ axis, and off the imaginary $p$ axis, are singularities on the next Riemann sheet.}
\label{fig:p-pole-plot10}
\end{figure}
This resurgent structure of singularities can also be seen in the Borel $p$ plane.  Fig. \ref{fig:p-pole-plot10} shows the separation of the Borel singularities in the $p$ plane after conformal mapping. Fig. \ref{fig:p-pole-plot10} displays the inverse conformal maps of the $z$ plane poles in  Fig. \ref{fig:z10}. With $N=50$ terms [red dots] we see clearly the accumulation of singularities at $p=\pm 2i$, in addition to an indication of singularities at $p=\pm 3i$. It is straightforward to generate more perturbative  input coefficients (i.e. larger $N$), to reveal even higher singularities. Contrast this with Fig. \ref{fig:pade-borel-poles}, showing the poles of the Pad\'e-Borel transform, before the conformal map, which shows no resurgent structure of repeated singularities.
\begin{figure}[htb]
\centering{\includegraphics[scale=.35]{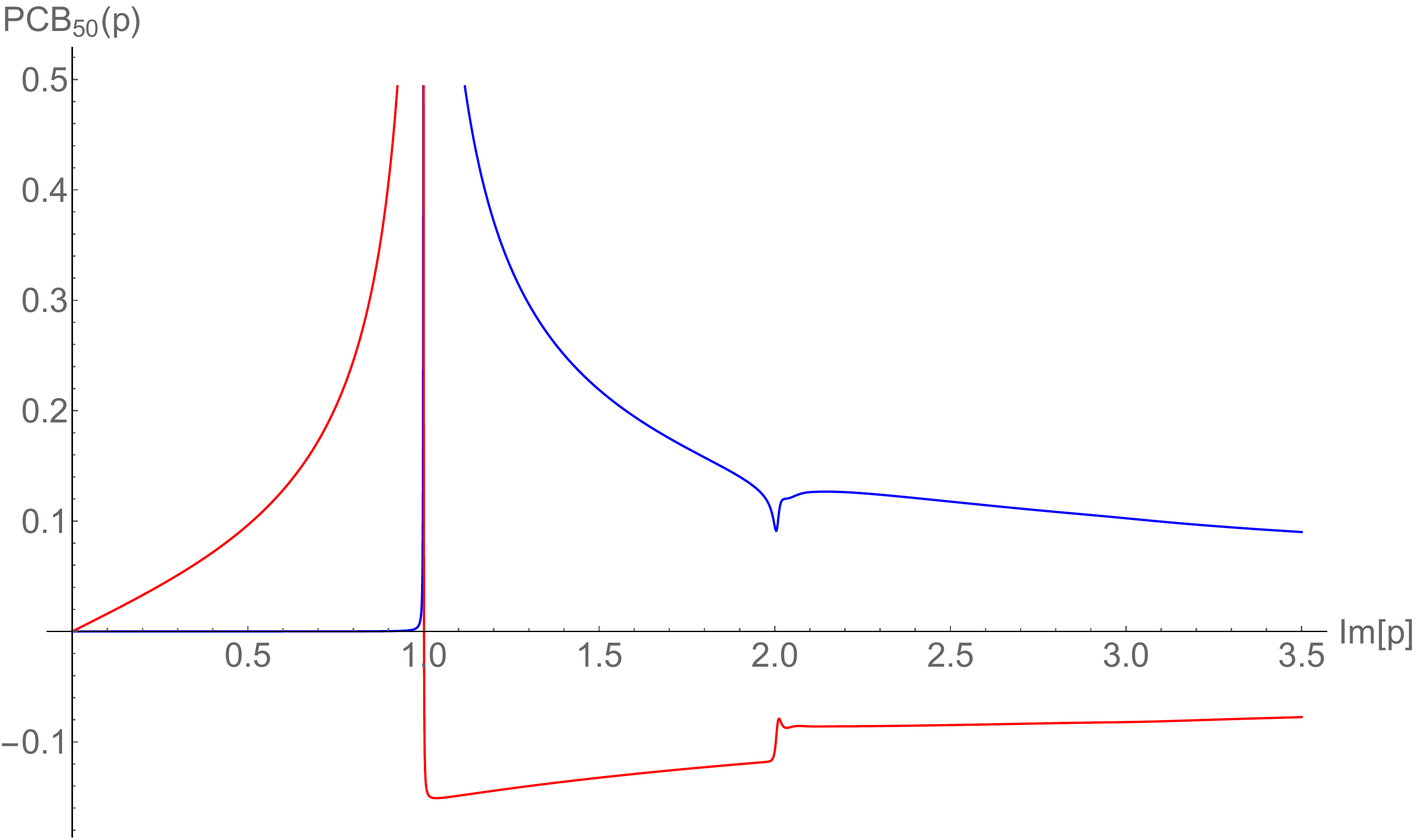}}
\caption{A plot of the real [blue curve] and imaginary [red curve] parts of  the conformally mapped Pad\'e-Conformal-Borel transform $\mathcal{PCB}_{50} [h](p)$, defined in (\ref{eq:pade-conformal-borel-p}),  along the imaginary $p$ axis  in the Borel $p$ plane. We see clearly the first two singularities at $p=+i$ and $p=+2i$, and an indication (after zooming in) of the third singularity at $p=+3i$ (this, and higher singularities, can be further resolved with higher $N$ values). A close-up view of the jump at $p=+2i$ is shown in Fig \ref{fig:jump-50}. }
\label{fig:borel-two-jumps-p50}
\end{figure}
\begin{figure}[htb]
\centering{\includegraphics[scale=.35]{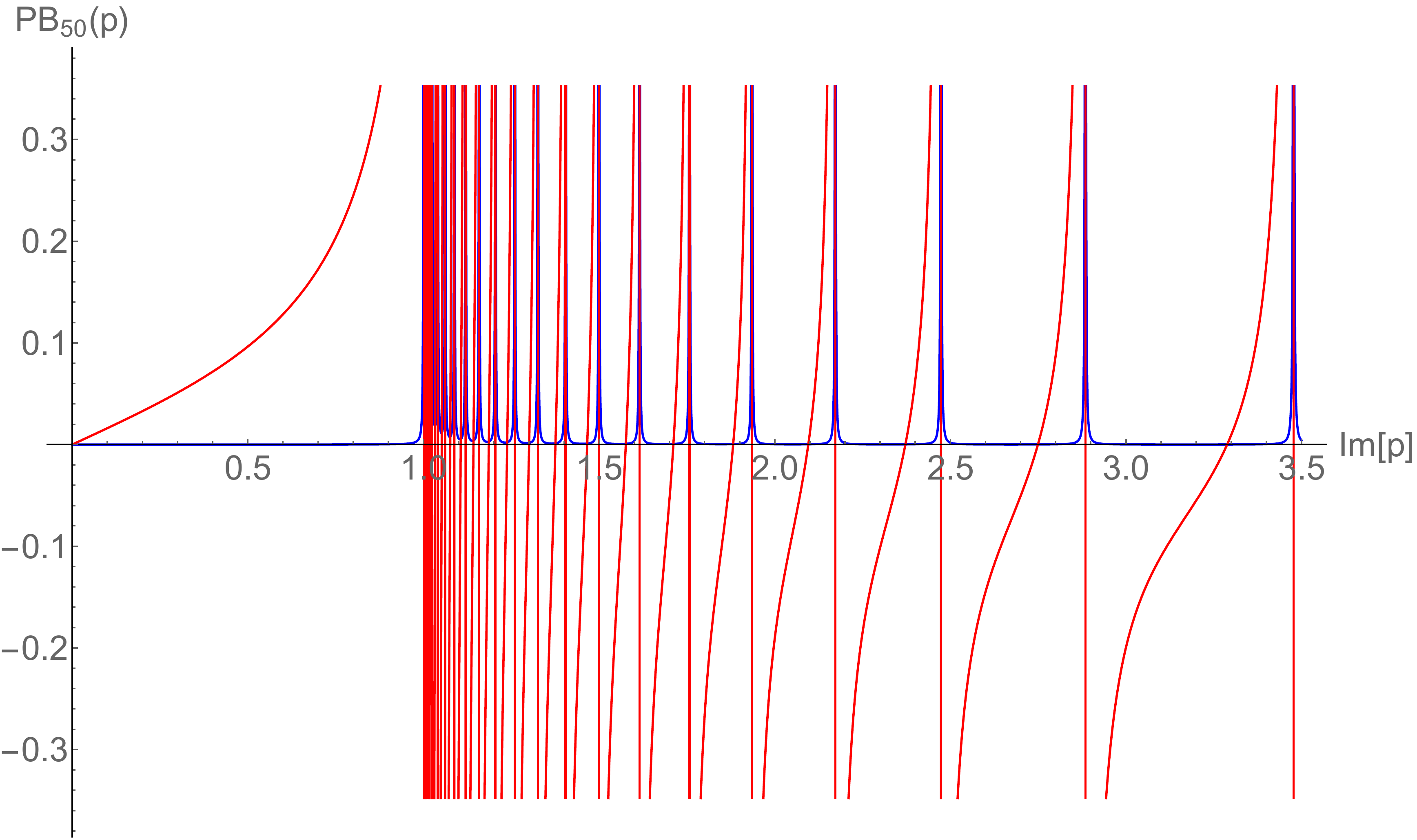}}
\caption{A plot of the real [blue curve] and imaginary [red curve] parts of the Pad\'e-Borel transform   $\mathcal{PB}_{50} [h](p)$, defined in (\ref{eq:pade-borel0}), along the imaginary $p$ axis in the Borel $p$ plane. Without the conformal map, none of the resurgent structure beyond the leading singularity can be resolved. Contrast with Fig. \ref{fig:borel-two-jumps-p50} and Fig. \ref{fig:jump-50} for the conformally mapped Borel function, where the resurgent singularity at $p=+2i$ is accurately resolved. }
\label{fig:pade-borel-no-jump}
\end{figure}

\begin{figure}[htb]
\centerline{\includegraphics[scale=0.35]{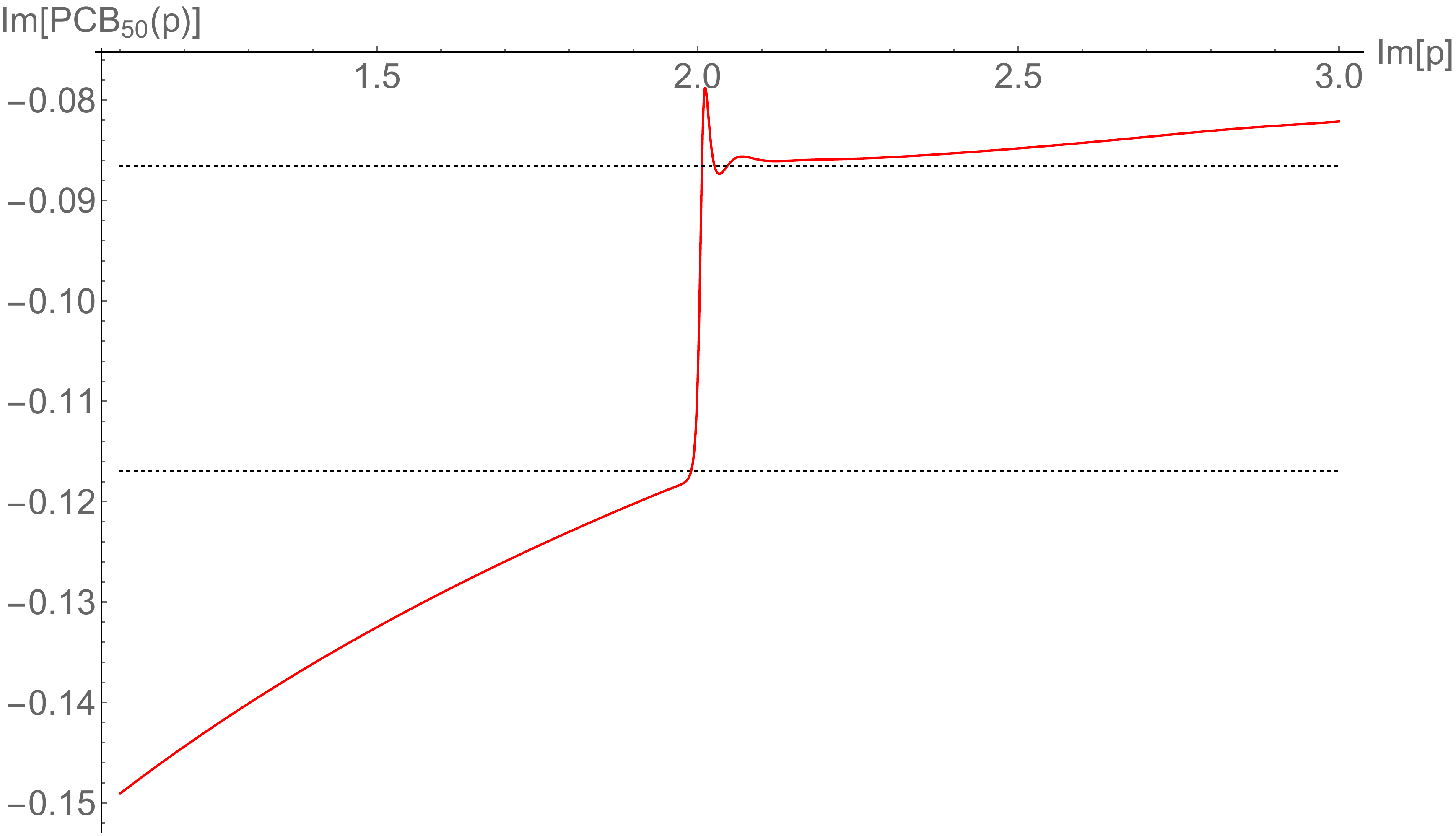}}
\caption{Plot of the imaginary part of the Pad\'e-Conformal-Borel transform ${\mathcal {PCB}}_{50}[h](p)$ along the imaginary $p$ line, zoomed-in from Fig. \ref{fig:borel-two-jumps-p50}, showing the jump at the second singularity $p=+2i$. The dotted horizontal lines correspond to the resurgent jump associated with a logarithmic singularity, of magnitude $\frac{1}{2}S^2$ as  in Eq. (\ref{eq:jump}).}
\label{fig:jump-50}
\end{figure}

This is not just a {\it qualitative} indication of the higher-order resurgent structure: the conformally mapped Borel transform also encodes {\it quantitative} information about the resurgent singularities. For example, comparing Figs. \ref{fig:pade-borel-stokes-10} and \ref{fig:conformal-pade-borel-stokes-10}, we see that the coefficient of the leading singularity can be resolved by the Pad\'e-Conformal-Borel function $\mathcal{PCB}_{10} [h](p)$, but not by the Pad\'e-Borel function $\mathcal{PB}_{10} [h](p)$. Turning to the higher resurgent singularities, at integer multiplies of $\pm i$, we contrast the behavior along the edge of the $p$ plane cut shown in Fig. \ref{fig:borel-two-jumps-p50} [for the Pad\'e-Conformal-Borel function $\mathcal{PCB}_{50} [h](p)$] with that shown in Fig. \ref{fig:pade-borel-no-jump} [for the Pad\'e-Borel function $\mathcal{PB}_{50} [h](p)$]. The Pad\'e-Conformal-Borel function resolves the first two singularities (with a hint of the third when zoomed-in), while the Pad\'e-Borel function does not resolve any of the higher resurgent singularities. Fig. \ref{fig:jump-50} shows a close-up of the imaginary part of the Pad\'e-Conformal-Borel function $\mathcal{PCB}_{N}[h](p)$ along the imaginary $p$ axis, in which we see clearly the resurgent jump at $p=+2i$. Furthermore, the {\it magnitude} of this jump coincides with that of the expected logarithmic behavior\footnote{The singularity is logarithmic because it is the inverse Laplace transform of the $\left(\frac{e^{-t}}{\sqrt{t}}\right)^2$ term \cite{costin-odes,costin-dmj}.} at the second singularity:
\begin{eqnarray}
\text{jump at second singularity}=\frac{1}{2}S^2 = \frac{1}{2}\left(\frac{1}{\pi}\sqrt{\frac{3}{5}}\right)^2
\approx 0.0304...
\label{eq:jump}
\end{eqnarray}

\section{Extrapolation into the Physical Complex $x$ Plane: The Origin and  the Pole Region}
\label{sec:re-expand}

\subsection{Re-expansion Method and Precision Extrapolation to the  Origin and Beyond}
\label{sec:origin}

The fact that the Pad\'e-Conformal-Borel transform ${\mathcal {PCB}}_{N}[h](p)$ encodes the resurgent structure of the Borel transform, even along the most difficult directions along the cuts on the imaginary axis in the Borel $p$ plane, suggests that the Laplace transform in (\ref{eq:num-y}) should be able to extrapolate the physical PI function $y(x)$ throughout a much larger region of the complex $x$ plane than is  possible with the Pad\'e-Borel transform. In this Section we explore this complex extrapolation, first all the way down to $x=0$, then onto the {\it negative} real $x$ axis, and then into the full complex $x$ plane. While this can be achieved by numerical contour integration, a much simpler and significantly more accurate method is the following two-step procedure.
\begin{enumerate}
\item
Use the Pad\'e-Conformal-Borel method to extrapolate from the asymptotic $x\to+\infty$ expansion (\ref{eq:largex}) down to some small value $x=x_0$ on the positive real axis, at which point the PI equation is satisfied to very high precision. As an illustration, we choose $x_0=3$. We extrapolate the function $h(t)$ from $t=+\infty$ along the real $t$ axis down to $t_0\equiv (24 x_0)^{5/4}/30\approx 7$, and evaluate $h(t_0)$ and $\dot{h}(t_0)$ by  straightforward numerical Borel integration:
\begin{eqnarray}
h_N(t_0)&=&\int_0^\infty dp\, e^{-p\, t_0}\, \mathcal{PCB}_{N}[h](p)
\label{eq:num-h1}\\
\dot{h}_N(t_0)&=&\int_0^\infty dp\,(-p) \, e^{-p\, t_0}\,\mathcal{PCB}_{N}[h](p)
\label{eq:num-h2}
\end{eqnarray}
Using the relation (\ref{eq:yh}) between $h(t)$ and $y(x)$, this yields extremely precise values for both $y(x_0)$ and $y^\prime(x_0)$. This precision is quantified below.

\item
Generate a Pad\'e approximant of the PI solution $y(x)$ in the physical $x$ plane, expanded about $x=x_0$. This Pad\'e approximant can be generated using the PI equation, expressing all higher derivatives $y^{(n)}(x_0)$ in terms of $y(x_0)$ and $y^\prime(x_0)$, in much the same way as the original perturbative coefficients were generated by expanding about $x=+\infty$. This step is algorithmic, but requires as input extremely precise values values of $y(x_0)$ and $y^\prime(x_0)$. But this is exactly what our first step of Pad\'e-Conformal-Borel extrapolation has produced.
\end{enumerate}

To quantify the precision of the first step, we measure how well the extrapolated function satisfies the PI equation. Analogous to (\ref{eq:num-h1})--(\ref{eq:num-h2}), we  compute $\ddot{h}_N(t_0)$ as a numerical integral, $\ddot{h}_N(t_0)=\int_0^\infty dp\,(-p)^2 \, e^{-p\, t_0}\,\mathcal{PCB}_{N}[h](p)$, which yields a value for  $y^{\prime\prime}(x_0)$. The precision  at $x_0$ is then measured by the degree to which the PI equation (\ref{eq:p1}) is satisfied.
For example, using the Pad\'e-Conformal-Borel transform $\mathcal{PCB}_{N}[h](p)$ to extrapolate down to $x_0=3$, starting with just $N=10$ terms of the asymptotic expansion at $x=+\infty$, we satisfy the PI equation at $x=x_0$ to 12  decimal places, and with $N=50$ terms it is satisfied to $29$ decimal places. Had we used instead the Pad\'e-Borel transform function $\mathcal{PB}_{N}[h](p)$, we still obtain impressive precision: 10 decimal places for $N=10$, and 22 decimal places for $N=50$:
\begin{eqnarray}
N=10\,\, \text{terms with Pad\'e-Conformal-Borel:} \quad \left[y''(x)-6y^2(x)+x\right]_{x=3}&=&O(10^{-13})\\
N=50\,\, \text{terms with Pad\'e-Conformal-Borel:} \quad \left[y''(x)-6y^2(x)+x\right]_{x=3}&=&O(10^{-30})\\
N=10\,\,\text{terms with Pad\'e-Borel:} \quad \left[y''(x)-6y^2(x)+x\right]_{x=3}&=&O(10^{-11})\\
N=50\,\, \text{terms with Pad\'e-Borel:} \quad \left[y''(x)-6y^2(x)+x\right]_{x=3}&=&O(10^{-23})
\end{eqnarray}
The precision increases as $\sqrt{N}$, in agreement with analytic estimates \cite{cd-2}.

This $x$ space Pad\'e approximation provides a simple analytic continuation of $y_N(x)$ into the complex plane. For example, we can evaluate $y_N(x)$ directly at the origin. We thereby obtain very precise values for the {\it tritronqu\'ee} initial conditions at the origin. With $N=10$ terms and  Pad\'e-Conformal-Borel input we obtain
\begin{eqnarray}
&&N=10\,\,\text{terms with Pad\'e-Conformal-Borel:} \nonumber\\
&&y(0)\approx  -0.187554308...  \nonumber\\
&&y^\prime(0)\approx  -0.304905560 ...  \nonumber\\
&&y^{\prime\prime}(0)\approx  0.211059715 ...\nonumber\\
&&\left[y''(x)-6y^2(x)+x\right]_{x=0}=O(10^{-9})
\label{eq:atzero-10}
\end{eqnarray}
and the PI equation is satisfied to 8 digit precision at the origin.
With $N=50$ terms and  Pad\'e-Conformal-Borel input we obtain
\begin{eqnarray}
&&N=50\,\, \text{terms with Pad\'e-Conformal-Borel:} \nonumber\\
&&y(0)\approx  -0.18755430834049489383868175759583299323116090976213899693337265167...\nonumber\\
&&y^\prime(0)\approx -0.30490556026122885653410412498848967640319991342112833650059344290... \nonumber\\
&&y^{\prime\prime}(0)\approx   0.21105971146248859499298968451861337073253247206264082468899143841... \nonumber\\
&&\left[y''(x)-6y^2(x)+x\right]_{x=0}=O(10^{-65})
\label{eq:atzero-50}
\end{eqnarray}
and the PI equation is satisfied to 64 digit precision at the origin. Pad\'e-Borel input also leads to precise values at the origin, but with lower precision. 
This level of precision should be compared with roughly 14 digits of precision for $y(0)$ and $y^\prime(0)$ obtained for Painlev\'e equations by the best current boundary-value, initial-value and Fredholm determinant numerical  methods that operate along the positive real axis \cite{fornberg,bornemann}. The ultimate reason for the remarkable improvement in precision is that our extrapolation method incorporates the resurgent structure of the function, which encodes global information about the function throughout the entire complex plane, not just along the positive real axis.

\begin{figure}[htb]
\centerline{\includegraphics[scale=.85]{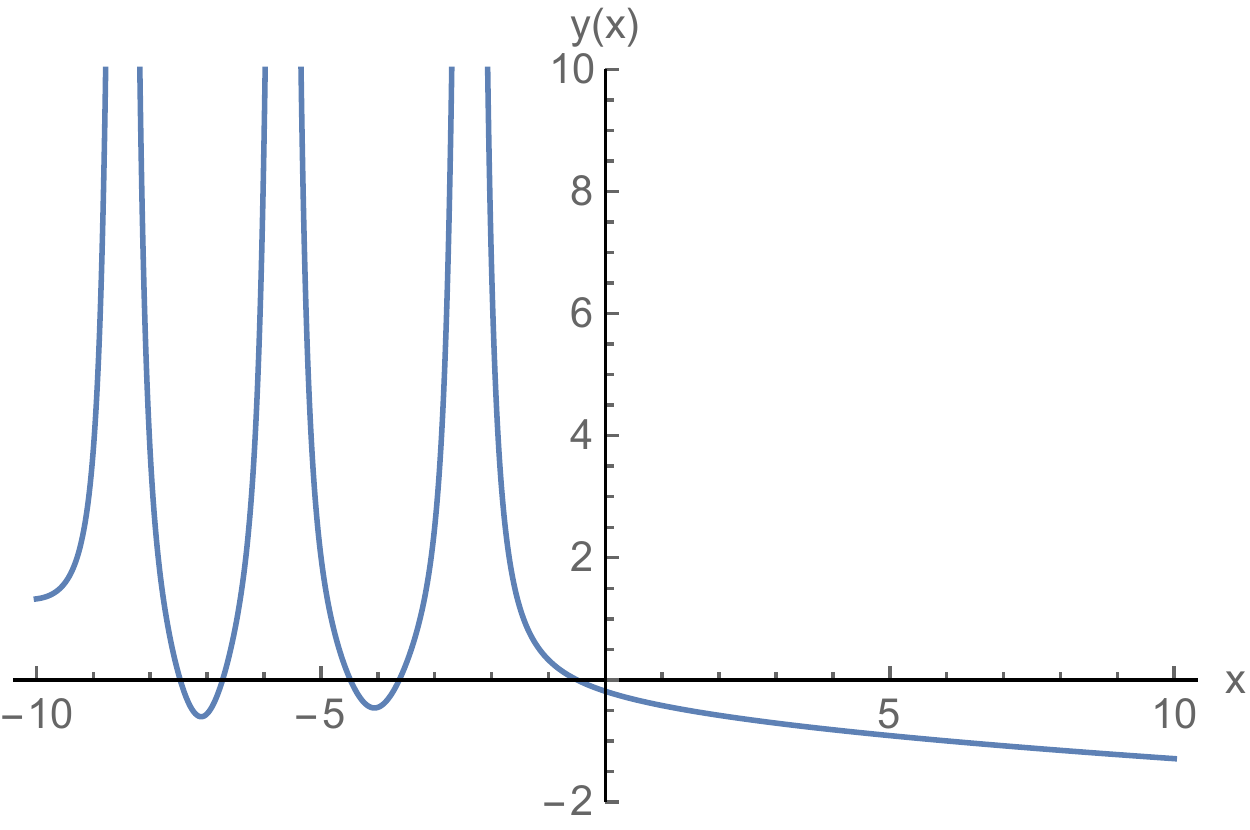}}
\caption{Plot of the extrapolated function $y(x)$ along the real $x$ axis, showing the first three poles on the negative real axis. The function $y(x)$ is obtained from $N=50$ input coefficients, and resolves accurately the first 3 poles on the negative real $x$ axis.}
\label{fig:negative-x-poles}
\end{figure}

As a further application, we can further continue the reconstructed function $y_N(x)$ along the {\it negative} real $x$ axis, where the PI 
{\it tritronqu\'ee} solution is known to have poles. Fig. \ref{fig:negative-x-poles} shows $y_N(x)$ in this region, with $N=50$ input coefficients, with the first three (double) poles clearly visible. Even with $N=10$  input coefficients we accurately resolve the first pole on the negative $x$ axis. The extrapolated function goes directly across the phase transition into this pole region, which we explore in further detail in the next section.

\subsection{Stokes Transition: Mapping the Tritronqu\'ee Pole Region}
\label{sec:poles}
\begin{figure}[htb]
\centerline{\includegraphics[scale=.7]{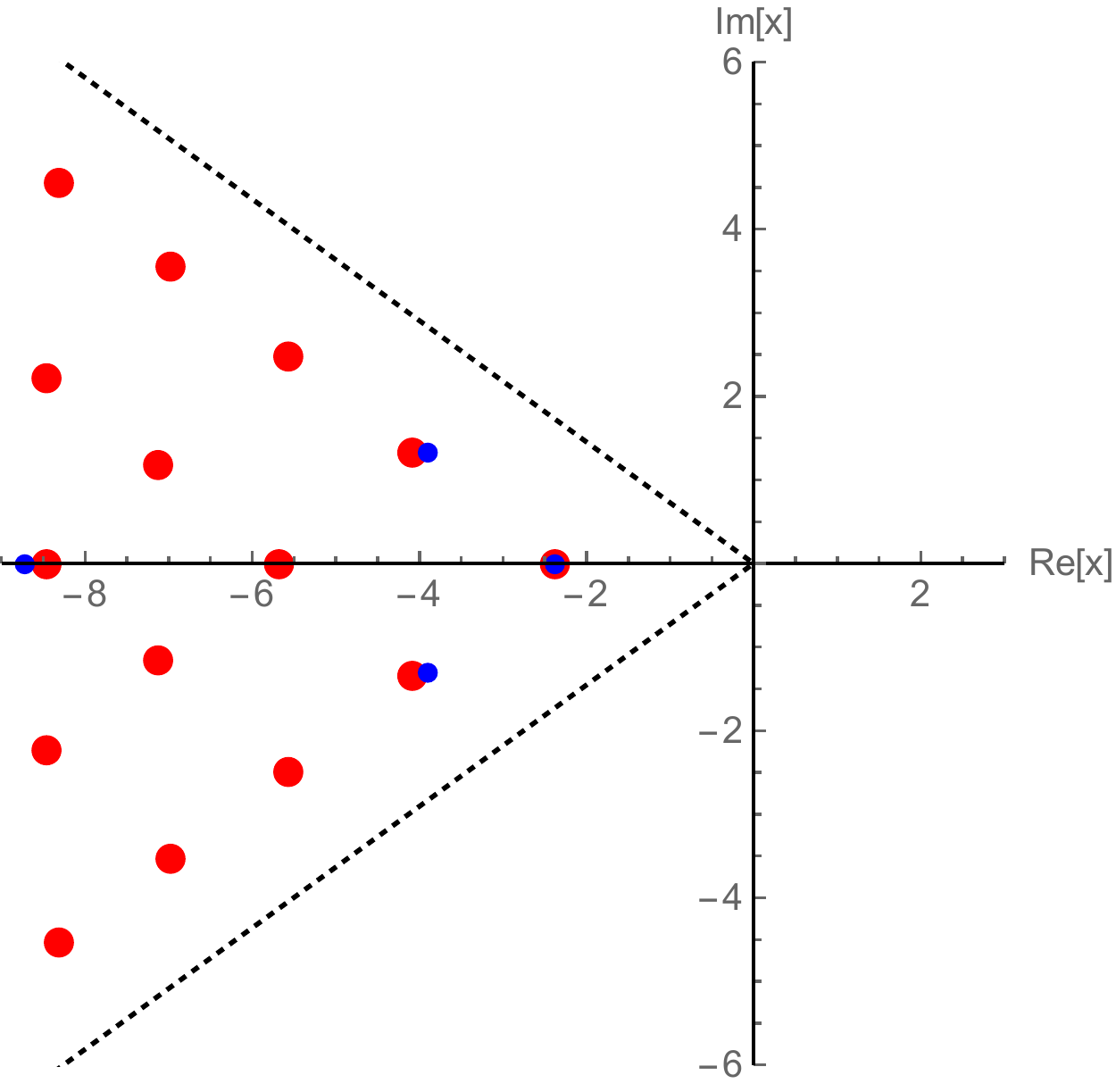}}
\caption{Poles of our extrapolated PI solution $y(x)$ in the complex plane. The smaller blue dots are extracted starting with just $N=10$ input coefficients (\ref{eq:input}) at $x\to+\infty$, while the larger red dots are extracted from $N=50$ input coefficients. The dashed lines mark the edges of the Stokes wedge for the {\it tritronqu\'ee}  pole region: $\frac{4\pi}{5}\leq {\rm arg}(x)\leq \frac{6\pi}{5}$. This confirms the Dubrovin conjecture \cite{dubrovin}, stating that for the {\it tritronqu\'ee} PI solution the poles only lie within this wedge region.}
\label{fig:wedge}
\end{figure}
\begin{figure}[htb]
\centerline{\includegraphics[scale=.7]{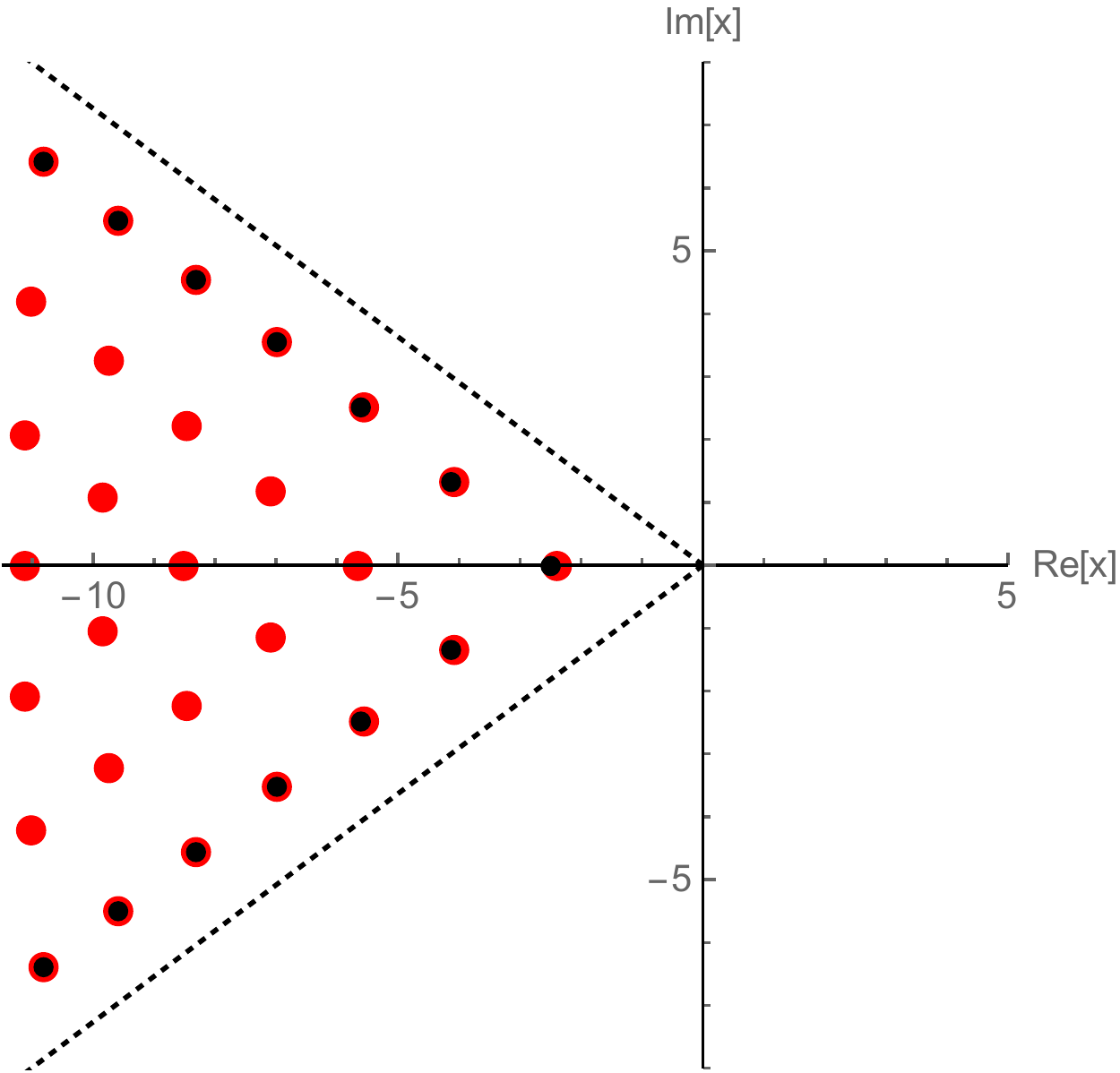}}
\caption{The first 28 {\it tritronqu\'ee} poles,  $x_n$ [red dots] of the Painlev\'e I solution $y(x)$, obtained from our Pad\'e-Conformal-Borel procedure, with $N=100$ input coefficients from (\ref{eq:input}), compared with the trans-asymptotic large $n$ expression $x_n^{\rm outer\,\,layer}$  in (\ref{eq:pole-approx1}) [smaller black dots]. The dashed lines mark the edges of the Stokes wedge for the {\it tritronqu\'ee}  pole region: $\frac{4\pi}{5}\leq {\rm arg}(x)\leq \frac{6\pi}{5}$. Note that even at $n=1$, the asymptotic large $n$ expression (\ref{eq:pole-approx1}) is surprisingly accurate.}
\label{fig:wedge2}
\end{figure}
In this Section we cross the Stokes transitions in the physical $x$ variable, at ${\rm arg}(x)=\pm \frac{4\pi}{5}$, and map the  {\it tritronqu\'ee} complex pole region. The Dubrovin conjecture \cite{dubrovin} states that for the {\it tritronqu\'ee} solution $y(x)$ to PI, which has the asymptotic expansion (\ref{eq:largex}) at $x=+\infty$ \cite{joshi,kitaev,kapaev}, the only poles lie in the wedge region: $4\pi/5 < {\rm arg}(x) < 6\pi/5$. This conjecture has been proved in \cite{costin-huang-tanveer}, and has been confirmed by several numerical analyses \cite{novokshenov,bornemann,fornberg}. Here we use our resurgent extrapolation to give another  high precision confirmation of this conjecture, combined with an analysis of the fine structure of the pole region. 

Fig. \ref{fig:wedge} shows the poles of our analytically continued solution $y_N(x)$, obtained by combining the Pad\'e-Conformal-Borel extrapolation with a Pad\'e expansion at $x_0=3$, using just $N=10$ [blue dots], or $N=50$ [red dots], coefficients as input data. We stress that given the $N$ input coefficients from (\ref{eq:input}), the rest of the computation is entirely algorithmic. We find it quite remarkable that with just 10 terms of the asymptotic expansion at $x\to+\infty$, the first three poles in the {\it tritronqu\'ee} pole region can be seen with reasonable precision. With 50 terms, we resolve the first 21 poles with a high degree of precision. It is a simple matter to work with even higher values of $N$, if further poles and/or higher precision are desired. See Fig. \ref{fig:wedge2}.

We see from Fig. \ref{fig:wedge}  that the poles do indeed lie within the expected $2\pi/5$ wedge, a numerical confirmation of the Dubrovin conjecture \cite{dubrovin}. To probe the {\it tritronqu\'ee} poles more precisely, we compare them with the locations predicted by the trans-asymptotic analysis of \cite{costin-costin-inventiones,costin-costin-huang}. An asymptotic expression for the outer layer of poles in the 
{\it tritronqu\'ee} wedge is \cite{costin-costin-inventiones,costin-costin-huang}:
\begin{eqnarray}
x_n^{\rm outer\,\,layer}\approx e^{\frac{2\pi\, i}{5}}\left(\frac{30^{4/5}}{24}\right) \left(2\pi\, i\, \left(n-\frac{1}{2}\right)+\ln\left(\frac{C}{12\sqrt{2\pi\, i\, n}}\right)-
\dots \right)^{4/5}\,\,, \quad n\to\infty 
\label{eq:pole-approx1}
\end{eqnarray}
and $(x_n^{\rm outer\,\,layer})^*$, 
where $C$ is proportional to the Stokes constant: $C=i \sqrt{\frac{6}{5 \pi }}$. 
In Fig. \ref{fig:wedge2} we compare the $n\to\infty$ trans-asymptotic estimate in (\ref{eq:pole-approx1}) with the numerical poles obtained from our Pad\'e-Conformal-Borel procedure, starting with $N=100$ coefficients at $x\to+\infty$. The agreement with the $n\to\infty$ asymptotics  is remarkable, even all the way down to $n=1$ for the pole closest to the origin. Asymptotic formulas for successive layers of lines of poles can be derived straightforwardly from recursion relations for adiabatic invariants \cite{costin-costin-inventiones,costin-costin-huang}.

We comment that in a very interesting paper \cite{novokshenov}, Novokshenov has made a Pad\'e expansion of the PI solution $y(x)$ directly at the origin in the physical $x$ variable, which can be ``tuned" to the {\it tritronqu\'ee} solution by adjusting the initial values $y(0)$ and $y^\prime(0)$ according to the requirement that the resulting Pad\'e approximant has no poles outside the {\it tritronqu\'ee} pole region wedge $4\pi/5 < {\rm arg}(x) < 6\pi/5$. To ensure that no poles leak outside of this region, this requires an extremely delicate search procedure, profoundly sensitive to the initial values $y(0)$ and $y^\prime(0)$; and therefore a very efficient Pad\'e algorithm was developed in \cite{novokshenov}. By contrast, in our $x$-space Pad\'e re-expansion method (see Sec. \ref{sec:origin}), the initial values $y(0)$ and $y^\prime(0)$ are automatically fixed by our high-precision Pad\'e-Conformal-Borel extrapolation, and no search is needed at all. 

\subsection{Stokes Wedges and {\it Tritronqu\'ee}  Connection Formulas}
\label{sec:tri-plots}

In this Section we plot our extrapolated PI solution $y(x)$, from Sec. \ref{sec:origin}, along the Stokes and anti-Stokes lines for the {\it tritronqu\'ee} solution. These are the most interesting and nontrivial directions in the complex $x$ plane, describing the transition boundaries between the five different Stokes wedges.
\begin{figure}[htb]
\centerline{\includegraphics[scale=.7]{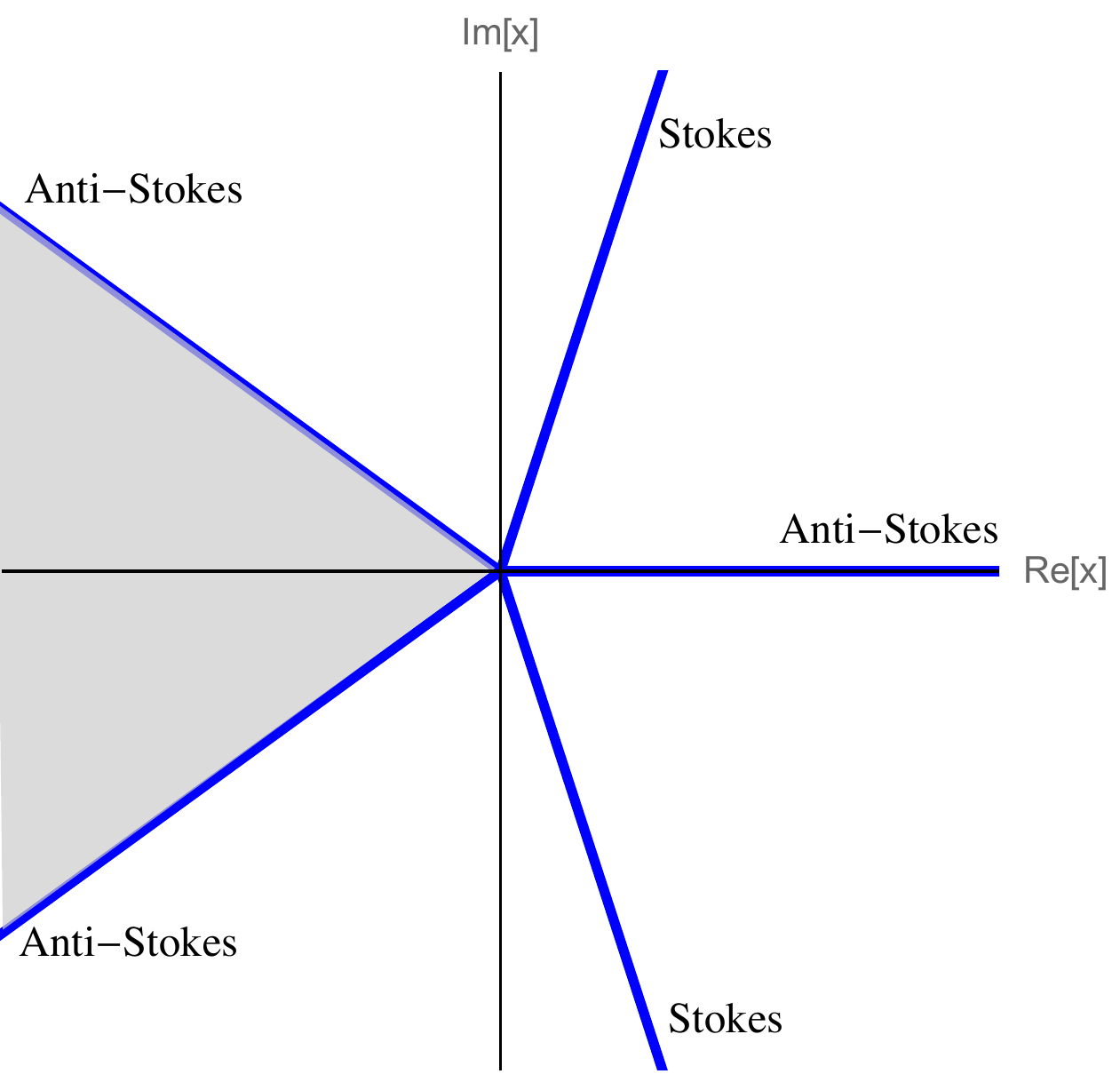}}
\caption{The five Stokes wedges of the PI equation (\ref{eq:p1}), showing the Stokes and anti-Stokes lines for the {\it tritronqu\'ee} solution, as lines with ${\rm arg}(x)=k\, \frac{2\,\pi}{5}$, with $k=0, \dots, 4$. The shaded  wedge is the {\it tritronqu\'ee}  pole region: $\frac{4\pi}{5}<\text{arg}(x)<\frac{6\pi}{5}$. Recall that our perturbative input (\ref{eq:input}), from which we have developed our extrapolation of the PI solution $y(x)$  throughout the complex $x$ plane, was obtained from the asymptotics as $x\to+\infty$ along the anti-Stokes line on the positive real $x$ axis. }
\label{fig:stokes-plot}
\end{figure}
Recall that the PI equation is invariant under rotation of the coordinate $x$ by $e^{\frac{2\pi\, i}{5}}$, and the function $y$ by $e^{\frac{4\pi\, i}{5}}$ \cite{boutroux,kapaev,kitaev,dubrovin,novokshenov,Garoufalidis:2010ya,joshi,takei}, and the asymptotics (\ref{eq:largex}) of the {\it tritronqu\'ee} solution defines the set of Stokes and anti-Stokes lines, ${\rm arg}(x)=k\, \frac{2\,\pi}{5}$ with $k=0, \dots, 4$,  as shown in Fig. \ref{fig:stokes-plot}.  Our perturbative input (\ref{eq:input}), from which we have developed our extrapolation of the PI solution $y(x)$  throughout the complex $x$ plane, was obtained from the large $x$ asymptotics along one particular direction:  the anti-Stokes line along the positive real axis. The behavior of $y(x)$ along the other Stokes and anti-Stokes lines is very different.  As a precise diagnostic check of the quality of our extrapolation into the complex $x$ plane, we can test the known analytic connection formulas of the PI  {\it tritronqu\'ee} solution, which relate the rotated {\it tritronqu\'ee} solutions along the Stokes and anti-Stokes lines \cite{boutroux,kapaev,kitaev,Garoufalidis:2010ya}.

\begin{figure}[htb]
\centerline{\includegraphics[scale=.9]{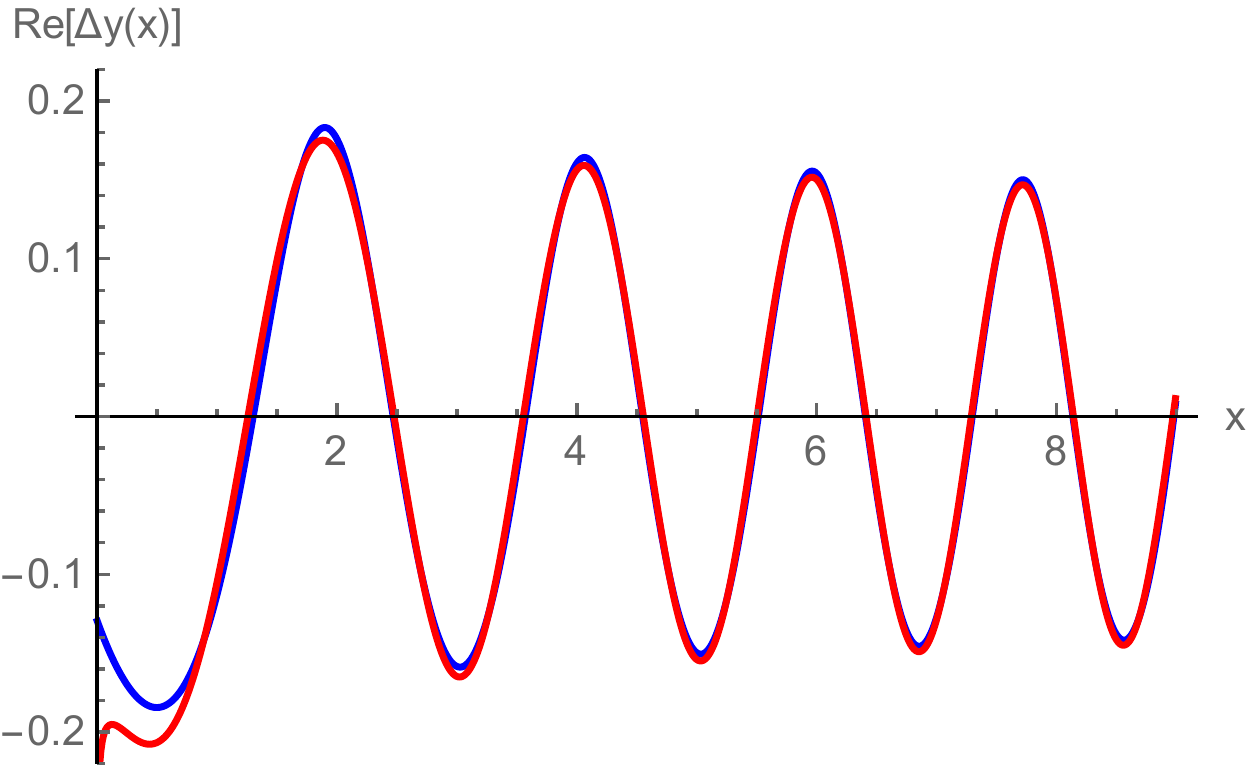}}
\caption{Plot of the {\it tritronqu\'ee} connection formula (\ref{eq:wedge1}). The blue curve is the real part of the left-hand-side of (\ref{eq:wedge1}), while the red curve is the real part of the right-hand-side, using our Painlev\'e I solution $y(x)$ extrapolated  into the complex plane.  Analogous plots can be made for the imaginary parts.}
\label{fig:wedge1-plot}
\end{figure}

\begin{figure}[htb]
\centerline{\includegraphics[scale=.9]{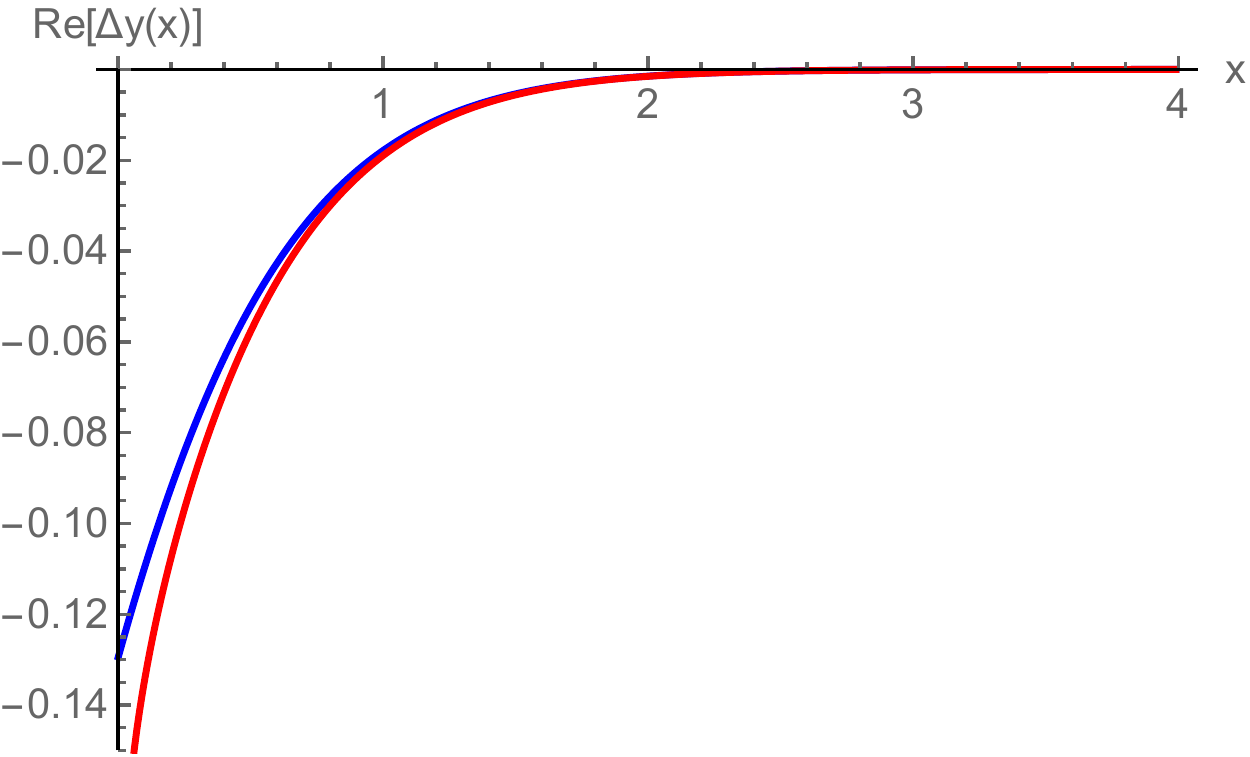}}
\caption{Plot of the {\it tritronqu\'ee} connection formula (\ref{eq:wedge2}). The blue curve is the real part of the left-hand-side of (\ref{eq:wedge1}), while the red curve is the real part of the right-hand-side, using our Painlev\'e I solution $y(x)$ extrapolated  into the complex plane. Analogous plots can be made for the imaginary parts.}
\label{fig:wedge2-plot}
\end{figure}

For example,  the {\it tritronqu\'ee} solution $y(x)$ along the positive $x$ axis, and along the anti-Stokes line ${\rm arg}(x)=\frac{4\pi}{5}$, at the edge of the pole region, are related by the following exact connection formula, exhibiting oscillatory behavior with a coefficient depending on the Stokes constant \cite{kapaev,kitaev,Garoufalidis:2010ya}:
\begin{eqnarray}
y(x)-e^{\frac{8\pi\, i}{5}}y\left(e^{\frac{4\pi\, i}{5}}\, x\right)\sim \frac{e^{-\frac{3\pi\, i}{4}}}{2^{5/4} 6^{1/8} \sqrt{\pi} }\frac{e^{-i (24\, x)^{5/4}/30}}{x^{1/8}}\qquad , \quad x\to \infty
\label{eq:wedge1}
\end{eqnarray}
Fig. \ref{fig:wedge1-plot} shows this combination using our extrapolated function $y(x)$, displaying the oscillatory behavior of (\ref{eq:wedge1}), capturing accurately both the period and the amplitude.
On the other hand, the {\it tritronqu\'ee} solution $y(x)$ along the Stokes  lines ${\rm arg}(x)=\pm \frac{2\pi}{5}$ are related by the following exact connection formula, exhibiting exponentially decaying behavior, also with a coefficient depending on the Stokes constant \cite{kapaev,kitaev,Garoufalidis:2010ya}:
\begin{eqnarray}
y(e^{-\frac{2\pi\, i}{5}}\, x)-e^{\frac{8\pi\, i}{5}}y\left(e^{\frac{2\pi\, i}{5}}\, x\right)\sim \frac{e^{-\frac{3\pi\, i}{4}+\frac{i\, \pi}{20}}}{2^{5/4} 6^{1/8} \sqrt{\pi} }\frac{e^{- (24\, x)^{5/4}/30}}{x^{1/8}}\qquad , \quad x\to \infty
\label{eq:wedge2}
\end{eqnarray}
Fig. \ref{fig:wedge2-plot} shows this combination using our extrapolated function $y(x)$, displaying the correct exponentially decaying behavior of (\ref{eq:wedge2}). 

\subsection{Fine Structure of the Tritronqu\'ee Poles}
\label{sec:pole-structure}

The {\it general} PI solution $y(x)$ is known to be meromorphic throughout the complex plane \cite{boutroux,gromak,fokas}, and to have poles throughout the complex plane, in all five Stokes wedges. Indeed, in the vicinity of a moveable pole, the {\it general} PI solution $y(x)$ has a Laurent expansion of the following form
\begin{eqnarray}
y(x)&\approx& \frac{1}{(x-x_{\rm pole})^2}+\frac{x_{\rm pole}}{10}(x-x_{\rm pole})^2+\frac{1}{6}(x-x_{\rm pole})^3+ h_{\rm pole} (x-x_{\rm pole})^4
 \nonumber\\
 &&+\frac{x_{\rm pole}^2}{300}(x-x_{\rm pole})^6 +\frac{x_{\rm pole}}{150}(x-x_{\rm pole})^7+\dots
\label{eq:general}
\end{eqnarray}
All further coefficients of this Laurent expansion are expressed as polynomials in the two parameters $x_{\rm pole}$ and $h_{\rm pole}$. Thus, any PI solution $y(x)$ is completely determined by two constants, $x_{\rm pole}$ and $h_{\rm pole}$, {\it in the vicinity of any one of its poles}.
The {\it tritronqu\'ee} is special in the sense that it has poles only in one Stokes wedge \cite{dubrovin,costin-huang-tanveer} (see Sec. \ref{sec:poles}). 
 Since our re-expansion method produces the most precise values for the pole closest to the origin, it is natural to use this first pole, $x_1$ and its associated expansion constant $h_1$, to characterize the {\it tritronqu\'ee} solution to PI. In fact, since there is nothing special about $x=0$ for the PI equation,  the data $(x_1, h_1)$ is a much more natural way to characterize the {\it tritronqu\'ee} solution than giving $y(0)$ and $y^\prime(0)$ at the origin. High precision values for the first pole and the associated expansion constant  are:
\begin{eqnarray}x_1&=&-2.3841687695688166392991458524493...
\nonumber\\
h_1&=&-0.062135739226177640896490141640...
\label{eq:poledata}
\end{eqnarray}
These agree with previously quoted values \cite{joshi}, although those values were given to significantly lower precision. 

To quantify the precision of the {\it tritronqu\'ee} values in (\ref{eq:poledata}), we expand our $N=50$ extrapolated $y(x)$ Pad\'e  function about its pole closest to the origin, $x_1$, and compare the terms of the expansion with the known general form (\ref{eq:general}): 
\begin{eqnarray}
y(x)&\approx&\frac{0.9999999999999999999999999999999999997886}{(x-x_1)^2}\nonumber\\
&&+3.5\times 10^{-35}-2.4\times 10^{-34} (x-x_1)\nonumber\\
&&-0.238416876956881663929914585244923803
   (x-x_1)^2 \nonumber\\
   &&+0.166666666666666666666666666666657864
   (x-x_1)^3 \nonumber\\
    &&-0.06213573922617764089649014164005140
   (x-x_1)^4\nonumber\\
   && +4\times 10^{-31} (x-x_1)^5 \nonumber\\
&&+0.0189475357392909503157755851627665
   (x-x_1)^6 \nonumber\\
&&-0.015894458463792110928660972350677
   (x-x_1)^7 +\dots
   \label{eq:first-pole-expansion}
   \end{eqnarray}
   We observe that the coefficient of the leading pole term equals 1 to 36 decimal places, and the next two terms vanish to 34 and 33 decimal places. 
The quadratic term coefficient agrees with $\frac{x_1}{10}$ to 31 decimal places, and the coefficient of the $(x-x_1)^3$ term equals $\frac{1}{6}$ to 31 decimal places. Furthermore, the coefficient of the $(x-x_1)^5$ term vanishes to 30 decimal places. Thus we estimate the precision of the constant $h_1$ as the coefficient of the $(x-x_1)^4$ term to 30 decimal places. As a further check, we note that the coefficients of the next two terms in the expansion agree with their exact values, $\frac{1}{300}x^2_1$ and $\frac{1}{150}x_1$, to 29 and 28 decimal places, respectively.
\begin{table}[htp]
\label{tab:xntable}
\begin{center}
\begin{tabular}{|c|c|}
\hline
pole label $n$ & pole location $x_n$\cr
\hline
1 &  -2.38416876956881663929914585244925489
\cr\hline 2 &
-4.07105552317228805393+1.33555121517567079952 i
  \cr \hline 3 &
 -4.07105552317228805393-1.33555121517567079952 i
 \cr\hline 4&
 -5.57356521477+2.48916297098 i
 \cr\hline 5&
 -5.57356521477-2.48916297098 i 
   \cr\hline 6&
-5.664602914
\cr\hline 
\end{tabular}
\caption{List of the first 6 {\it tritronqu\'ee} poles $x_{\rm pole}$ from equation (\ref{eq:general}), starting with $N=50$ input coefficients. The displayed digits correspond to our estimate of the precision of our computation near each pole: see text.}
\end{center}
\end{table}%

\begin{table}[htp]
\begin{center}
\begin{tabular}{|c|c|}
\hline
pole label $n$ & second expansion constant $h_n$\cr
\hline
1 & -0.0621357392261776408964901416401 
\cr\hline 2
   &-0.1491925267759824-0.0650559915206451 i
   \cr\hline 3 
   &-0.1491925267759824+0.0650559915206451 i 
   \cr\hline 4 
   & -0.2485278-0.1390038 i 
   \cr\hline 5
   & -0.2485278+0.1390038 i 
   \cr\hline 6 
   & -0.238327
   \cr\hline 
\end{tabular}
\caption{List of the second expansion constant $h_{\rm pole}$ from equation (\ref{eq:general}), for the first 6 {\it tritronqu\'ee} poles, starting with $N=50$ input coefficients. The displayed digits correspond to our estimate of the precision of our computation near each pole: see text.}
\end{center}
\label{tab:hntable}
\end{table}%
 In Tables I and II we record our results for the first 6 poles, $x_n$, and the associated expansion constants, $h_n$, obtained from our numerical extrapolation, based on $N=50$ input coefficients. The number of digits shown for each pole is determined by the method described above for $x_1$, applied to each pole $x_n$. The precision degrades  quickly for the poles further from the origin, but this can be improved by taking larger $N$, and also by combining with trans-asymptotic estimates such as (\ref{eq:pole-approx1}), which become much more precise for the poles further from the origin. In this paper we have not implemented these refinements, but we quote these initial values because of their relevance to the quantum mechanical spectral problem for cubic oscillators, and because only very low order values exist in the literature for the first two real poles, $x_1$ and $x_6$ \cite{masoero}. A more detailed numerical study of the {\it tritronqu\'ee} pole values is left for future work.
This is motivated by results connecting poles of PI solutions to spectral properties of cubic oscillators \cite{masoero,bender,novokshenov2}, in analogy to results relating pole behavior of Painlev\'e III solutions with the spectrum of the Mathieu equation \cite{novokshenov-mathieu,lukyanov,gorsky}, and pole behavior of Painlev\'e VI solutions to spectral properties of an associated Heun equation \cite{litvinov,gd}.

\section{Conclusions}

We have studied the numerical extrapolation of the solution $y(x)$ to the Painlev\'e I equation (\ref{eq:p1}), starting from a  {\it finite}  number of terms in the asymptotic expansion at $x\to+\infty$. Combining standard methods of Borel transforms, Pad\'e approximants, and conformal mapping, together with aspects of resurgent asymptotics, we obtain a surprisingly precise extrapolation throughout the complex $x$ plane. Our initial asymptotic expansion (\ref{eq:largex}) generates the {\it tritronqu\'ee} solution to PI, and we tested the precision of our extrapolation by comparing it with known analytic properties such as exact connection formulae and trans-asymptotic pole expressions. The extrapolated function crosses smoothly across the non-linear Stokes transitions into the pole region. Both Pad\'e-Borel and the conformally mapped Pad\'e-Conformal-Borel extrapolations produce high quality extrapolations, but the latter is more accurate  in a larger range of the complex plane. The ultimate reason for this is that the conformal map resolves more efficiently the underlying regularity of the resurgent structure in the Borel plane. The resurgent extrapolation method is very general,  not relying on the integrability of the Painlev\'e I equation, and should be applicable to the much broader class of resurgent problems in physics. Several extensions and refinements of the resurgent extrapolation method are possible, which may become more relevant in other problems where fewer perturbative coefficients can be generated, and/or if these coefficients are generated with  limited precision. We list some of these refinements here, and further details will be given in subsequent papers.

{\it Duality Bootstrap:} the expansion of $y(x)$ as $x\to+\infty$ maps directly to the $p\to 0$ behavior of the Borel transform, and correspondingly the behavior of $y(x)$ as $x\to 0$ maps directly to the $p\to \infty$ behavior of the Borel transform. In certain cases the problem may be one of {\it interpolation}, in which some knowledge about both $x\to+\infty$ and $x\to 0$ is known, but one wants to interpolate between these two expansions in a way that accurately describes intermediate values of $x$. In this case, the duality between large and small $x$, and small and large $p$ can be used to develop an iterative "bootstrap" procedure. In the case of PI, our extrapolation method already produced sufficient numerical precision, but in other problems this duality bootstrap can be a powerful additional tool.

{\it Singularity Tuning:} as mentioned in Sec. \ref{sec:res-poles}, the leading Borel singularity for PI is of square root character [see (\ref{eq:nature})], and is mapped to a pole by the conformal map (\ref{eq:map1}). In other problems, where the leading singularity is not a square-root branch point, one can apply ramified re-definitions of the expansion parameter and Borel variable in order to engineer a leading pole after conformal mapping. This facilitates the resurgent  separation of the singularities, resulting in improved precision.

{\it Higher Riemann sheets:} we observed numerically that the Pad\'e-Conformal-Borel produces singularities on higher Riemann sheets. These singularities could be used for further higher precision tests of resurgence, and also for a systematic numerical investigation of \'Ecalle's medianization \cite{ecalle,sauzin}.

{\it Continued Fractions, Orthogonal Polynomials and Pad\'e Approximants:} there is a deep connection between Pad\'e, continued fractions, orthogonal polynomials and conformal mapping, which we discuss in detail in \cite{cd-2}, but we comment briefly on the basic ideas here. The outcome of this connection is that one can derive analytic estimates of the precision that can be obtained by Pad\'e-Borel and Pad\'e-Conformal-Borel with $N$ input coefficients. Resurgence appears in these extrapolations due to the fact that given $N$ terms of an expansion, Pad\'e can predict the next term with exponential precision. Analytic continued fractions \cite{wall}, in one of the most useful normalizations,  are rational functions of the form 
\begin{eqnarray}
R_N(p)=b_0+\frac{m_0(p)}{\displaystyle 1+\frac{m_1(p)}{1+\cdots}}=:[b_0,[m_0(p),1],[m_1(x),1]\cdots[m_N(p),1]]
\label{eq:cf}
\end{eqnarray}
where $m_k(p)$ are monomials, $m_k(p)=c_k \, p^{q_k}$. Given a power series, the constants $c_k$ and $q_k$  are uniquely determined by requiring that the Maclaurin series of $R_N$ coincides with the given series to the highest possible order allowed by the total degree of $R_N$. It follows that $R_N(p)$ is in fact a way of rewriting a near-diagonal Pad\'e approximant $P_N(p)/Q_N(p)$, deg($P_N$)$\approx$deg($Q_N$). This is an important representation of Pad\'e approximants for a number of reasons, including the fact that the $c_k$ are much smaller than the Pad\'e coefficients,  and that  the $c_k$ often have asymptotic expansions at large  $k$, which give analytic information about the series. For 
Painlev\'e I we calculated $R(p)=[0,[\frac{4p}{25},1],[\frac{49p^2}{75},1],\cdots]$, and observed empirically that, for large $k$, $m_k(p)\sim \frac14 p^2$. A straightforward induction argument shows that the polynomials $P_N$ and $Q_N$ satisfy  two-step recurrence relations  \cite{wall}, which can therefore be associated with orthogonal polynomials. Then the asymptotics of $m_k(p)$  follows from  Szeg\"o asymptotics of orthogonal polynomials \cite{szego}. We can then estimate the successive error as $R_N-R_{N-1}\sim A(p)\left(z(p)\right)^{2N}$, for some $A(p)$ independent of $N$, where the conformally mapped variable $z(p)$ from (\ref{eq:map1}) arises from solving the asymptotic recursion formulas. 
A deep result of Damanik and Simon characterizes the asymptotics $m_k=p^2/4+\epsilon_k$, with $\epsilon_k$ small enough \cite{damanik}. This can be used to show that the accuracy in $\mathbb{C}$  of a Pad\'e-Borel approximant with $N$ terms is $O((1-z^2(p))^{-1}z(p)^{2N})$, roughly the same as that of the conformally mapped {\em Taylor series} in the unit disk. In particular, if $|p|\ge 1$ and $\Re\,p=0$, then $|z(p)|=1$; and thus standard  Pad\'e-Borel approximants diverge everywhere on the cuts. On the other hand, for small $p$, $z^2(p)\sim p^2/4$ meaning that, well inside the unit disk, Pad\'e-Borel gives an $O(4^{-N})$ accuracy improvement over the Taylor series. And, away from the cuts, for large $p$ and $N$,  a standard Borel-Pad\'e approximation has accuracy $O(p e^{-2N/|p|})$, which translates to  $O(N^{3/4}|x|^{-5/4}e^{-\sqrt{8 N |x|}})$ accuracy in the physical domain $x$. And the improvement of the conformally mapped Pad\'e-Conformal-Borel transform in (\ref{eq:pade-conformal-borel-p}) over the Pad\'e-Borel transform in (\ref{eq:pade-borel0}) also scales like $4^{-N}$. Further details and applications will appear in \cite{cd-2}.

{\it Resurgent Numerical Analysis:} our results suggest that it may also be fruitful to incorporate some ideas of resurgence into the sophisticated numerical analysis methods of \cite{bornemann,fornberg}.

\bigskip

This material is based upon work supported by the U.S. Department of Energy, Office of Science, Office of High Energy Physics under Award Number DE-SC0010339 (GD), and by the National Science Foundation under Award Number DMS 1515755 (OC). Both authors thank  the KITP at Santa Barbara for its hospitality during the Fall 2017 program 
``Resurgent Asymptotics in Physics and Mathematics'' where much of this work was begun. 
Research at KITP is supported by the National Science Foundation under
Grant No. NSF PHY-1125915. 
We thank A. Its, P. Nevai, and  C. Bender  for helpful discussions, comments and correspondence.

\end{document}